\documentclass{amsart}
\usepackage{amsmath}
\usepackage{amsfonts}
\usepackage{graphics}
\usepackage{epsfig}
\usepackage{amssymb}
\usepackage{amscd}

\newtheorem{thm}{Theorem}[section]
\newtheorem{cor}[thm]{Corollary}

\newtheorem{prop}[thm]{Proposition}
\theoremstyle{definition}
\newtheorem{Def}[thm]{Definition}
\newtheorem*{ack}{Acknowledgement}
\theoremstyle{remark}
\newtheorem*{rem}{Remark}
\newtheorem{Case}{Case}
\newtheorem{step}{Step}

\theoremstyle{definition}
\newtheorem{ex}{Example}[section]

\numberwithin{equation}{section}
\numberwithin{figure}{section}

\def\exp{{\text{\rm{exp}}}}

\def\rchi{{\hbox{\raise1.5pt\hbox{$\chi$}}}}
\def\Aut{{\text{\rm{Aut}}}}

\def\isom{\cong}
\def\tensor{\otimes}

\def\Affine{{\text{\rm{Affine}}}}
\def\Ker{{\text{\rm{Ker}}}}

\def\ie{\hbox{\it i.e. }}
\begin{document}

\title[Ribbon Graphs, Quadratic Differentials,
 and Curves over $\overline{\mathbb{Q}}$]
{Ribbon Graphs, Quadratic Differentials
on Riemann Surfaces, and
Algebraic Curves Defined over $\overline{\mathbb{Q}}$}
\author[M.~Mulase]{Motohico Mulase}
\address{
Department of Mathematics\\
University of California\\
Davis, CA 95616--8633}
\email{mulase@math.ucdavis.edu}
\author[M.~Penkava]{Michael Penkava}
\address{
Department of Mathematics\\
University of Wisconsin\\
Eau Claire, WI 54702--4004}
\email{penkavmr@uwec.edu}
\date{September 28, 1998}

\allowdisplaybreaks
\setcounter{section}{-1}
\begin{abstract}
It is well known that there is a bijective correspondence 
between metric ribbon graphs and compact Riemann surfaces 
with meromorphic Strebel differentials. In this article,
it is  proved that Grothendieck's correspondence between
\emph{dessins d'enfants} and Belyi morphisms is 
a special case of this correspondence. For a
metric ribbon graph with edge length $1$,
an algebraic curve over $\overline{\mathbb{Q}}$ and
a Strebel differential on it is constructed. It is also
shown   that the
critical trajectories of the measured foliation that
is determined by the Strebel differential recover the 
original metric ribbon graph. 
Conversely, for every Belyi morphism, a unique 
Strebel differential is constructed such that 
the critical
leaves of the measured foliation it determines form 
a metric ribbon graph of edge length $1$, which coincides
with the 
corresponding dessin d'enfant.  
\end{abstract}
\maketitle
\tableofcontents
\section{Introduction}\label{sect: intro}
In this article we give a
self-contained explanation of the relation
between ribbon graphs (combinatorial data),
 algebraic curves
defined over $\overline{\mathbb{Q}}$ (algebraic
and arithmetic data),
and Strebel differentials on
Riemann surfaces (analytic data).


For a given Riemann surface, we ask when  it has
the structure of an algebraic curve
defined over the field $\overline{\mathbb{Q}}$ of
algebraic numbers. A theorem of Belyi \cite{Belyi}
answers this question
saying  that a nonsingular Riemann surface
 is an algebraic curve defined over
$\overline{\mathbb{Q}}$
 if and only if there is a holomorphic
map of the Riemann surface onto $\mathbb{P}^1$
that is ramified only at $0$, $1$ and $\infty$.
Such a map is called a \emph{Belyi map}.

Grothendieck discovered that
there is a natural bijection between the set of
isomorphism classes of
connected ribbon graphs and the set of
isomorphism classes of Belyi maps. Thus a ribbon
graph defines a Riemann surface with a complex
structure and, moreover, its algebraic structure over
$\overline{\mathbb{Q}}$. If we start with a Belyi map,
then the corresponding ribbon graph is just the
inverse image of the interval $[0,1]$ of
$\mathbb{P}^1$. Grothendieck called these
graphs \emph{Child's Drawings}
 (\emph{dessins d'enfants}). We refer to
 \cite{Schneps} for more detail on this subject.

Another correspondence between ribbon graphs and
Riemann surfaces, this time between metric ribbon
graphs and arbitrary Riemann surfaces,
has been known since the work of Harer, Mumford,
Penner, Thurston, and others
(see \cite{Harer}). In this second correspondence,
a ribbon graph arises as the union of critical
leaves of a measured foliation defined on a
Riemann surface by a
meromorphic quadratic differential
called a \emph{Strebel differential}
\cite{Gardiner,Strebel}.
When the Riemann surface is defined over
$\overline{\mathbb{Q}}$, it coincides with the same surface that is given
by the Grothendieck correspondence between ribbon graphs and
algebraic curves defined over
$\overline{\mathbb{Q}}$.

In this paper we give constructive proofs of these facts
using \emph{canonical coordinate systems} arising from
Strebel differentials on a
Riemann surface. The Child's Drawings,
Belyi maps and Strebel differentials are
related in a very simple way, and they are explicitly
described in terms of  simple formulas.

Although these formulas could
be written down in a few pages (see Section~\ref{sect: belyi}),
for the sake of completeness we
have included a detailed description of the theory which relates
metric ribbon graphs and moduli spaces of Riemann surfaces with
marked points.

In Section~\ref{sect: ribbon}, we give a definition of
ribbon graphs and their automorphisms.
Thurston's \emph{orbifolds}
and their Euler characteristics are defined in
Section~\ref{sect: orbifold}. With these preparations,
in Section~\ref{sect: metric}
we prove that the space of all isomorphism
classes of metric ribbon graphs (i.e., ribbon
graphs with a positive real number assigned to
each edge) is a differentiable orbifold. Since a
simplicial complex can be arbitrarily singular,
this statement is not trivial.
In Section~\ref{sect: strebel}, we review Strebel
differentials on a Riemann surface, and construct
a canonical coordinate system. The natural bijection
between the space of metric ribbon graphs and
the moduli space of  Riemann surfaces with marked
points is given in Section~\ref{sect: combi} by means of an
explicit construction of the Strebel differential in terms of
canonical coordinates corresponding to a metric ribbon graph.
Finally, in Section~\ref{sect: belyi}, we
give an explicit formula for the Belyi map
corresponding to  an arbitrary ribbon graph
in terms of these canonical local coordinates.

The correspondence between metric ribbon graphs, quadratic
differentials and the
moduli space of Riemann surfaces is well-known to
specialists. Moreover, formulas for these correspondences have
appeared in the literature, in more or less explicit form.
Nevertheless, the authors believe that our formulation
of this correspondence is more precise, and
leads to a simple formulation of some
properties of algebraic curves defined over
$\overline{\mathbb{Q}}$,
which may be useful for further study.

\begin{ack}
The authors thank Bill Thurston for explaining
his work \cite{STT} to them. They are also
grateful to Francesco Bottacin and Regina Parsons who has
made valuable
suggestions and improvements to the article.
The work is partially supported by funding from the
University of California, Davis, the University of
Wisconsin, Eau Claire, and the NSF.
\end{ack}

\section{Ribbon graphs}
\label{sect: ribbon}

A graph is a finite collection of points and line
segments connected in certain ways, and
a ribbon graph is a graph drawn on an oriented
surface. A more careful definition of these
objects is necessary when we consider their
\emph{isomorphism classes}.

\begin{Def}\label{def: graph}
A
\emph{graph} $\Gamma = ({\mathcal{V}}, {\mathcal{E}}, i)$
consists of a
finite set ${\mathcal{V}}
= \{V_1, V_2, \cdots, V_v\}$ of \emph{vertices} and a finite set
${\mathcal{E}}$ of
\emph{edges},
together with a map $i$ from ${\mathcal{E}}$ to the set
$({\mathcal{V}}\times {\mathcal{V}})/\mathfrak{S}_2$ of
unordered pairs of vertices,
called the \emph{incidence relation}. An edge and a vertex
are said to be incident if the vertex is in the image of
the edge under $i$.
The quantity
$$
a_{jk} = |i^{-1}(V_j, V_k)|
$$
gives the number of edges that connect two vertices
$V_j$ and $V_k$.
The \emph{degree}, or
\emph{valence},  of a vertex $V_j$ is the number
$$
\deg(V_j) = \sum_{k\ne j} a_{jk} + 2a_{jj},
$$
which is the number of edges incident to the vertex. A loop,
that is, an edge with
only one incident vertex, contributes twice to the
degree of its incident vertex.
The degree of every vertex is
required to be positive (no isolated vertices).
The \emph{degree sequence} of $\Gamma$ is the
ordered list of degrees of the
vertices:
$$
(\deg(V_1), \deg(V_2), \cdots, \deg(V_v)),
$$
where the vertices are arranged so that the
degree sequence is non-decreasing.
\end{Def}

In this
article, for the most part, we  shall consider only graphs
whose  vertices  all have degree
at least $3$.

\begin{figure}[htb]
\begin{center}
\epsfig{file=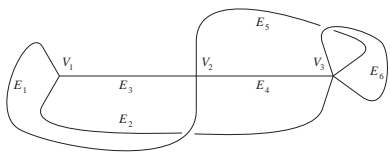, width=2in}
\end{center}
\caption{A graph consisting of $3$ vertices
and $6$ edges.}
\label{fig: graph}
\end{figure}

\begin{Def}
\label{def: traditional graph isomorphism}
A \emph{ traditional graph isomorphism}
$\phi = (\alpha, \beta)$ from a graph
$\Gamma = ({\mathcal{V}}, {\mathcal{E}}, i)$
to another graph $\Gamma' = ({\mathcal{V}}',
{\mathcal{E}}', i')$
is a pair of bijective maps
$$
\alpha: {\mathcal{V}}\overset{\sim}\longrightarrow
{\mathcal{V}}' \qquad\text{and}
\qquad \beta:{\mathcal{E}}\overset{\sim}
\longrightarrow {\mathcal{E}}'
$$
that preserve the incidence relation, i.e.,
a diagram
\begin{equation}
\begin{CD}
{\mathcal{E}} @>{i}>> ({\mathcal{V}}\times {\mathcal{V}})/
\mathfrak{S}_2\\
@V{\beta}V{\wr}V @V{\wr}V{\alpha\times \alpha}V\\
{\mathcal{E}}' @>>{i'}> ({\mathcal{V}}'\times {\mathcal{V}}')/
\mathfrak{S}_2
\end{CD}
\end{equation}
commutes.
\end{Def}

The traditional graph automorphism is not
the natural notion when we consider graphs
in the context of Riemann surfaces and
Feynman diagram expansions. The reason
is that the above definition of a graph
does not distinguish between the half-edges,
so that one cannot distinguish which vertex
is associated to which half-edge, and thus the group of
traditional group automorphisms is smaller than the automorphism
group we will need to consider. One can treat the notion of
graphs with distinguished half edges independently,
but it is possible to embed the theory of such
graphs within the ordinary definition of graphs
by introducing the notion of the \emph{edge refinement}
of a graph
$\Gamma = (\mathcal{V},\mathcal{E},i)$,
which is the graph
$$
\Gamma_{\mathcal{E}} = ({\mathcal{V}}\coprod
{\mathcal{V}}_{\mathcal{E}}, {\mathcal{E}
\coprod\mathcal{E}}, i_{\mathcal{E}})
$$
 with the middle point of each edge of
$\Gamma$
added as a degree 2 vertex, where
 ${\mathcal{V}}_{\mathcal{E}}$ denotes the
set of all these midpoints
of edges.
The set of vertices of $\Gamma_{\mathcal{E}}$ is
the disjoint union ${\mathcal{V}}\coprod
{\mathcal{V}}_{\mathcal{E}}$, and the set of
edges is the disjoint union
 ${\mathcal{E}}\coprod {\mathcal{E}}$ because
the midpoint $V_E$ divides the edge $E$ into two
parts. The incidence relation is described by
a map
\begin{equation}\label{eq: incidence for edge refinement}
i_{\mathcal{E}}:
{\mathcal{E}}\coprod {\mathcal{E}}\longrightarrow
{\mathcal{V}}\times {\mathcal{V}}_{\mathcal{E}},
\end{equation}
because each edge of $\Gamma_{\mathcal{E}}$ connects
exactly one vertex of ${\mathcal{V}}
$ to a vertex of ${\mathcal{V}}_{\mathcal{E}}$. An
edge of $\Gamma_{\mathcal{E}}$ is called a
\emph{half-edge}
of $\Gamma$. For every vertex $V\in \mathcal{V}$ of
$\Gamma$, the set $i_{\mathcal{E}} ^{-1}(\{V\}\times
\mathcal{V}_{\mathcal{E}})$
consists  of half-edges incident to $V$. Note that
 we have
$$
\deg(V) = |i_{\mathcal{E}} ^{-1}(\{V\}\times
{\mathcal{V}}_{\mathcal{E}})|.
$$

\begin{figure}[htb]
\begin{center}
\epsfig{file=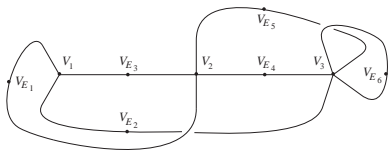, width=2in}
\end{center}
\caption{Edge refinement of the graph of
Figure~\ref{fig: graph}, with $9$ vertices
and $12$ edges.}
\label{fig: edge refinement}
\end{figure}

\begin{Def}
\label{def: graph automorphism}
Let $\Gamma = ({\mathcal{V}}, {\mathcal{E}}, i)$
be a graph without vertices of degree less than $3$.
The \emph{automorphism group}
$\Aut(\Gamma)$ is the group of
 traditional graph automorphisms
of the edge refinement $\Gamma_{\mathcal{E}}$.
\end{Def}

\begin{figure}[htb]
\begin{center}
\epsfig{file=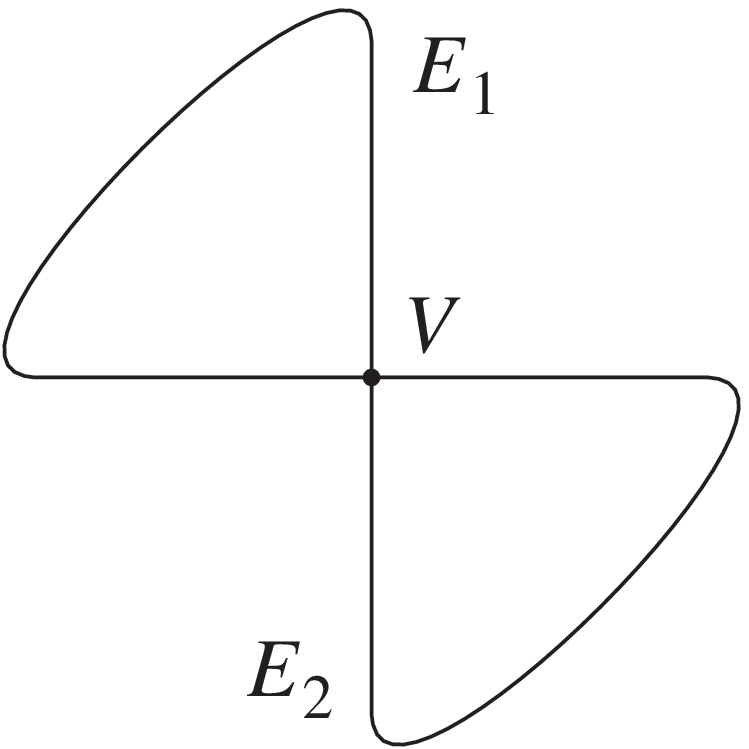, width=0.5in}
\hskip0.5in
\epsfig{file=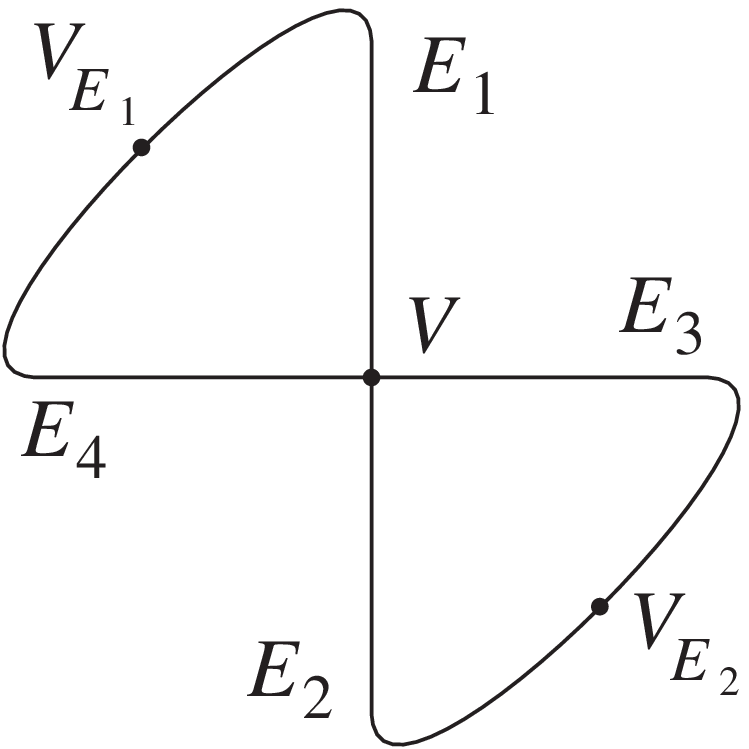, width=0.5in}
\end{center}
\caption{A graph and its edge refinement.}
\label{fig: graph automorphism}
\end{figure}

For example, the graph with
one degree $4$ vertex and two edges has
$(\mathbb{Z}/2\mathbb{Z})^3$ as its automorphism group,
while the traditional graph  automorphism group  is
just $\mathbb{Z}/2\mathbb{Z}$
(Figure~\ref{fig: graph automorphism}).

\begin{Def}
\label{def: connectedness}
Let $\Gamma$ be a graph. Two edges $E_1$ and
$E_2$ are \emph{connected} if there is a vertex
$V$ of $\Gamma$ such that both $E_1$ and $E_2$
are incident to $V$. A \emph{sequence of connected
edges} is an ordered set
\begin{equation}
\label{eq: sequence of connected edges}
(E_1,E_2,\cdots,E_k)
\end{equation}
of edges of $\Gamma$ such that $E_j$
and $E_{j+1}$ are connected for every
$j=1, 2, \cdots, k-1$. A graph $\Gamma$
is \emph{connected} if for every pair of
vertices $V$ and $V'$ of $\Gamma$, there is
a sequence of connected edges
(\ref{eq: sequence of connected edges})
for some integer $k$
such that $E_1$ is incident to $V$ and $E_k$ is
incident to $V'$.
\end{Def}

In this article we consider only connected
graphs.

\begin{Def}\label{def: ribbon graph}
A \emph{ribbon graph} (or \emph{fatgraph})
 is a graph $\Gamma
  = ({\mathcal{V}},{\mathcal{E}},i)$ together
with  a cyclic ordering
on the set of half-edges incident to each vertex of $\Gamma$.
\end{Def}

\begin{figure}[htb]
\begin{center}
\epsfig{file=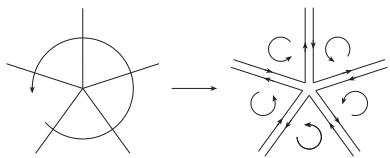, width=2in}
\end{center}
\caption{Oriented strips determined by the cyclic order at a vertex.}
\label{fig: cross road}
\end{figure}

A ribbon graph can be represented on a positively oriented plane
(\ie a plane with counter-clockwise orientation)
as a set of points corresponding to the vertices, connected by
arcs for each edge between the points corresponding to its incident vertices,
arranged so that the cyclic order of edges at a
vertex corresponds to the orientation of the plane.
Intersections of the arcs at points other than the vertices are
ignored. The half edges incident to a vertex can be replaced
with thin strips joined at the vertex,
with the cyclic order at the vertex determining
a direction on the boundaries of the strip
(Figure~\ref{fig: cross road}).

\begin{figure}[htb]
\begin{center}
\epsfig{file=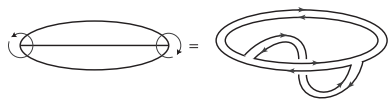, width=2in}
\end{center}
\caption{Oriented surface with boundary determined by a ribbon graph.}
\label{fig: cyclic order and ribbon graph}
\end{figure}

The strips corresponding to the two half edges are connected
following the orientation of their boundaries to form ribbons,
determining a figure which is no longer planar, but is an oriented surface with
boundary given by the
boundaries of the ribbons
(Figure~\ref{fig: cyclic order and ribbon graph}).

\begin{figure}[htb]
\begin{center}
\epsfig{file=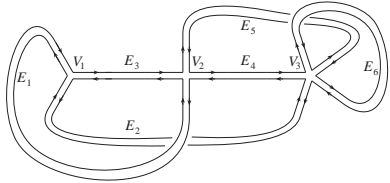, width=2in}
\end{center}
\caption{Ribbon graph of Figure~\ref{fig: graph} as an oriented surface.}
\label{fig: ribbon graph}
\end{figure}

Thus a ribbon graph can be considered as
an oriented surface with boundary, as is illustrated by
Figure~\ref{fig: ribbon graph}.

\begin{Def}
\label{def: ribbon graph automorphism}
An automorphism of a ribbon graph
 $\Gamma $ is an automorphism of the underlying
graph that preserves the cyclic ordering of
half-edges at every vertex.
\end{Def}
 Since we  deal  mainly with ribbon
graphs from now on, we use the notation
$\Aut(\Gamma)$ for  the
automorphism group of a ribbon graph $\Gamma$.
The characteristic difference between a  graph and
a ribbon graph is that the latter has a \emph{boundary}.

\begin{Def}\label{def: boundary of a ribbon graph}
Let $\Gamma  = ({\mathcal{V}},{\mathcal{E}},i,c)$
be a ribbon graph,
where $c$ denotes the cyclic ordering of half-edges
at each vertex. A \emph{directed edge} $\overrightarrow{E}$
is an ordering
$E^-$ and $E^+$
of the half-edges that form the edge $E$.
%
A \emph{boundary component}(hole) of $\Gamma $ is a
cyclic sequence of connected directed
 edges
$$
(\overrightarrow{E_0},
\overrightarrow{E_1},  \cdots,
\overrightarrow{ E_{q-1}},
\overrightarrow{E_q} =
\overrightarrow{E_0})
$$
such that
the half-edges $E_i ^+$ and $E_{i+1} ^-$ are incident to
a vertex
$V_i$ of $\Gamma$, with
$E_i ^+$ immediately preceding
$E_{i+1} ^-$
with respect
to the cyclic order assigned to the half-edges at
$V_i$.
\end{Def}

The ribbon graphs of
Figure~\ref{fig: cyclic order and ribbon graph}
and
Figure~\ref{fig: ribbon graph}
have only one boundary component.
We denote by $b(\Gamma )$ the
\emph{number of
boundary components} of a ribbon graph
$\Gamma $.

\begin{Def}
\label{def: automorphism fixing boundary components}
The group of ribbon graph
automorphisms of $\Gamma $ that
preserve the boundary components is denoted by
$\Aut_{\partial}(\Gamma )$, which is a
subgroup of
$\Aut(\Gamma )$.
\end{Def}

Since a boundary component of a ribbon graph is
defined to be a cyclic sequence of
directed edges,
the topological realization of the ribbon graph
has a well-defined orientation and
each boundary component has an induced orientation
that is compatible with the cyclic order. Thus we can
attach an oriented disk to each boundary component
of a ribbon graph $\Gamma $ so that the total
space, which we denote by $C(\Gamma)$, is
a compact oriented topological surface.

The attached
disks and the underlying graph
$\underline{\Gamma}$ of a
ribbon graph $\Gamma $ defines a
cell-decomposition of $C(\Gamma )$. Let
$v(\Gamma)$ denote the number of  vertices and
$e(\Gamma)$ the number of edges of
 $\Gamma$. Then the genus $g(C(\Gamma ))$
of the closed surface $C(\Gamma )$
is determined by a  formula for the Euler characteristic:
\begin{equation}
\label{eq: formula for Euler characteristic of a surface}
v(\Gamma) - e(\Gamma) + b(\Gamma )
= 2 - 2g(C(\Gamma )).
\end{equation}

The ribbon graph of
Figure~\ref{fig: cyclic order and ribbon graph}
has two vertices, three edges and one boundary
component. Thus the surface $C(\Gamma)$ is
a torus.

\begin{figure}[htb]
\begin{center}
\epsfig{file=ribbonfig/figtorus.eps, width=1in}
\end{center}
\caption{Cell-decomposition of a torus by
a graph drawn on it.}
\label{fig: cell-decomposition of a torus}
\end{figure}

The ribbon graph of Figure~\ref{fig: ribbon graph}
has three vertices, six edges and one boundary
component. Thus the genus of the closed surface
$C(\Gamma)$ associated with this ribbon graph is 2.

\begin{figure}[htb]
\begin{center}
\epsfig{file=ribbonfig/figg2graph.eps, width=1.5in}
\end{center}
\caption{Cell-decomposition of a surface of
genus $2$ by a ribbon graph.}
\label{fig: genus2 graph}
\end{figure}

In the next section we  study
\emph{metric ribbon graphs}, which are
ribbon graphs with a \emph{metric}, that is, an assignment
of a positive real number (length) to
each edge of the graph.
The set of all metrics on $\Gamma$
determines a topological space
homeomorphic to $\mathbb{R}^{e(\Gamma)}_+$, on which
the automorphism group of the ribbon graph acts.
We wish to
study the  structure of the space of isomorphism
classes of metrics
under this
action. A graph automorphism can act trivially
on the space of metrics on $\Gamma$, and this happens precisely when the
automorphism preserves the edges of $\Gamma$, but possibly
interchanges some of its half-edges.
Let us determine
all ribbon graphs that have
a non-trivial graph automorphism
acting trivially on the set of edges.

\begin{Def}\label{def: exceptional graphs}
A ribbon graph $\Gamma$ is \emph{exceptional} if
the natural homomorphism
\begin{equation}
\label{eq: hom auto to edges}
\phi_{\Gamma} : \Aut(\Gamma)
\longrightarrow \mathfrak{S}_{e(\Gamma)}
\end{equation}
of the automorphism group of $\Gamma$ to
the permutation group of edges is \emph{not}
injective.
\end{Def}

Let $\Gamma$ be an exceptional graph
and $\sigma\in \Ker(\phi_\Gamma)$ a nontrivial
automorphism of $\Gamma$.
Since none of the
edges are interchanged by $\sigma$,  the graph
can have at most two vertices. If the graph has two
vertices, then $\sigma$ interchanges
the vertices while all edges are fixed. The only
possibility is a  graph  with two vertices
of degree $j$, $(j\ge 3)$,
as in Figure~\ref{fig: cyclic order and ribbon graph} and
in Figure~\ref{fig: exception1} below.

\begin{figure}[htb]
\centerline{\epsfig{file=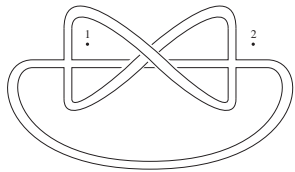, width=1.5in}}
\caption{Exceptional graph type 1.}
\label{fig: exception1}
\end{figure}

If $j$ is odd, then it has only one boundary component,
as in Figure~\ref{fig: cyclic order and ribbon graph}.
The genus of the surface
$C(\Gamma )$ is
given by
\begin{equation}\label{eq: jodd}
g(C(\Gamma )) = \frac{j-1}{2}.
\end{equation}
For an even $j$, the graph has two boundary components,
as in Figure~\ref{fig: exception1}, and
\begin{equation}\label{eq: jeven}
g(C(\Gamma )) = \frac{j-2}{2}.
\end{equation}
In both cases, the automorphism group is
the product group
\begin{equation}\label{eq: exceptional automorphism 1}
\Aut(\Gamma) = \mathbb{Z}/2\mathbb{Z} \times
\mathbb{Z}/j\mathbb{Z} ,
\end{equation}
with the factor $\mathbb{Z}/2\mathbb{Z}$
acting trivially on the set of edges.

When $\Gamma$ has two boundary components,
then
$$
\Aut_{\partial}(\Gamma) = \mathbb{Z}/j\mathbb{Z},
$$
which is a factor of
(\ref{eq: exceptional automorphism 1}). Note that
$\Aut_{\partial}(\Gamma)$ acts faithfully on the
set of edges in this case. Of course, in  the case
of one boundary component,
$\Aut_{\partial}(\Gamma)$ coincides with $\Aut(\Gamma)$, so
it does not act faithfully.

\begin{figure}[htb]
\centerline{\epsfig{file=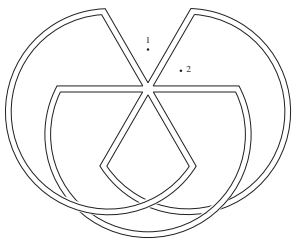, width=1.5in}}
\caption{Exceptional graph type 2}
\label{fig: exception2}
\end{figure}

To obtain the one-vertex case, we only need to contract one
of the edges of the two-vertex case considered
above.  The result is a  graph with
one vertex of degree $2k$,
 as shown in Figure~\ref{fig: exception2}.

When $k$ is even, the graph has only one boundary
component,
and the genus of the surface $C(\Gamma )$ is
\begin{equation}\label{eq: keven}
g(C(\Gamma )) = \frac{k}{2} .
\end{equation}
If $k$ is odd, then the graph has two boundary
components and the genus is
\begin{equation}\label{eq: kodd}
g(C(\Gamma )) = \frac{k-1}{2} .
\end{equation}
The automorphism group is $\mathbb{Z}/(2k)\mathbb{Z}$,
but the action on $\mathbb{R}_+ ^{k}$ factors
through
$$
\mathbb{Z}/(2k)\mathbb{Z}\longrightarrow
\mathbb{Z}/k\mathbb{Z} .
$$
Here again, in the 2 component case the automorphism group fixing the
boundary,
$\Aut_{\partial}(\Gamma)=
\mathbb{Z}/k\mathbb{Z}$,  acts faithfully
on the set of edges.

We have  thus classified all exceptional graphs.
These exceptional graphs
 appear  for arbitrary
genus $g$.
The graph of Figure~\ref{fig: exception1}
 has two distinct labelings of the boundary
components, but since they can be interchanged by
the action of a ribbon graph automorphism, there is
only one equivalence class of
ribbon graphs with labeled boundary over this underlying
ribbon graph. The automorphism group that preserves
the boundary  is $\mathbb{Z}/4\mathbb{Z}$. Thus the
space of  metric ribbon graphs with labeled boundary is
$\mathbb{R}_+ ^4/(\mathbb{Z}/4\mathbb{Z})$.
The change of labeling, or the action of $\mathfrak{S}_2$,
has a non-trivial effect on the graph level, but does
not act at all on the space
$\mathbb{R}_+ ^4/(\mathbb{Z}/4\mathbb{Z})$.
The space of metric ribbon graphs is also
$\mathbb{R}_+ ^4/(\mathbb{Z}/4\mathbb{Z})$,
which is  not the $\mathfrak{S}_2$-quotient of
the space of  metric ribbon graphs with labeled
boundary.

The other example of an exceptional graph,
Figure~\ref{fig: exception2}, gives another interesting
case. This time  the space of  metric
ribbon graphs with labeled
boundary and the space of  metric ribbon
graphs without referring to the boundary are both
$\mathbb{R}_+ ^3/(\mathbb{Z}/3\mathbb{Z})$. The
group $\mathfrak{S}_2$ of changing the labels
on the boundary has again
no effect on the space.

The analysis of exceptional graphs shows that
labeling all edges does not induce labeling of
the boundary components of a ribbon graph.
However, if we label all half-edges of a ribbon
graph, then we  have a labeling of the boundary
components as well. We will come back to this
point when we study the orbifold covering of
the space of metric ribbon graphs by the space of
metric ribbon graphs with labeled boundary components.
\section{Orbifolds and the Euler Characteristic}
\label{sect: orbifold}

A  space obtained by  patching
pieces of the form
$$
\frac{\text{smooth open ball}}{\text{finite group}}
$$
together
was called a \emph{$V$-manifold} by Satake
\cite{Satake} and an
\emph{orbifold} by Thurston \cite{Thurston}. {F}rom
the latter
we cite:

\begin{Def}\label{def: definition of orbifold}
An \emph{orbifold} $Q=\left(X(Q), \{U_i\}_{i\in I}, \{G_i\}_{i\in I},
\{\phi_i\}_{i\in I}\right)$ is
a set of data consisting of
\begin{enumerate}
\item a Hausdorff topological
space $X(Q)$ that is  called the \emph{underlying space},
\item a
locally finite open covering
$$
X(Q) = \bigcup_{i\in I} U_i
$$
of  the underlying space,
\item a set of homeomorphisms
$$
\phi_i : U_i \overset{\sim}{\longrightarrow}
\widetilde{U}_i/G_i .
$$
where
$\widetilde{U}_i$ is an open subset of $\mathbb{R}^n$
and $G_i$ is a finite group acting faithfully on $\widetilde{U}_i$.
\end{enumerate}
Whenever $U_i\subset U_j$, there is an injective
group homomorphism
$$
f_{ij} : G_i \longrightarrow G_j
$$
and an embedding
$$
\tilde\phi _{ij} : \widetilde{U}_i \longrightarrow
\widetilde{U}_j
$$
such that
$$
\tilde\phi _{ij}(\gamma x) = f_{ij}(\gamma)
\tilde\phi _{ij}(x)
$$
for every $\gamma\in G_i$ and $x\in \widetilde{U}_i$,
and such that the diagram below commutes.
\begin{equation*}
\begin{CD}
\widetilde{U}_i @>{\tilde\phi _{ij}}>> \widetilde{U}_j\\
@VVV @VVV\\
\widetilde{U}_i/G_i @>{\phi_{ij} = \tilde\phi _{ij}/G_i}>>
\widetilde{U}_j/f_{ij}(G_i)\\
@| @VVV\\
\widetilde{U}_i/G_i @>>> \widetilde{U}_j/G_j\\
@A{\phi_i}A{\wr}A @A{\wr}A{\phi_j}A\\
U_i @>{\text{inclusion}}>>U_j .
\end{CD}
\end{equation*}
\end{Def}

The space $Q$ is called an orbifold
\emph{locally modeled on $\mathbb{R}^n$ modulo
finite groups}.
An orbifold is said to be
\emph{differentiable}
if the group $G_i$ is a finite subgroup of  the
orthogonal group $O(n)$ acting on $\mathbb{R}^n$,
and the local models $\mathbb{R}^n/G_i$ are glued
together by  diffeomorphisms.

\begin{Def}\label{def: orbifold covering}
A surjective map
$$
\pi : Q_0 \longrightarrow Q_1
$$
of an orbifold $Q_0$ onto $Q_1$ is said to be
an \emph{orbifold covering} if the following
conditions are satisfied:
\begin{enumerate}
\item The map $\pi$ induces a surjective continuous map
$$
\pi : X(Q_0)\longrightarrow X(Q_1)
$$
between the underlying spaces, \emph{which is not generally
a covering map of the topological spaces}.
\item For every $x\in Q_0$, there is an open
neighborhood $U\subset Q_0$, an open
subset $\widetilde{U}\subset \mathbb{R}^n$,
a finite group $G_1$ a subgroup
  $G_0\subset G_1$, and homeomorphisms
$
U \overset{\sim}{\longrightarrow} \widetilde{U}/G_0
$
and
$
\pi(U) \overset{\sim}{\longrightarrow} \widetilde{U}/G_1
$
such that the diagram below commutes.
\begin{equation*}
\begin{CD}
U @>{\pi}>> \pi(U)\\
@V{\wr}V V @VV{\wr}V\\
\widetilde{U}/G_0 @>>> \widetilde{U}/G_1
\end{CD}
\end{equation*}
\item For every  $y\in Q_1$,
there is an open neighborhood $V$ of $y$,
an open subset $\widetilde{V} \subset \mathbb{R}^n$,
a finite group $G'_1$, a subgroup $G'_0\subset G'_1$,
and a connected component $U'$ of $\pi^{-1}(V)$
making the diagram below commute.
\begin{equation*}
\begin{CD}
U' @>{\pi}>> V\\
@V{\wr}VV @VV{\wr}V\\
\widetilde{V}/G'_0 @>>> \widetilde{V}/G'_1
\end{CD}
\end{equation*}
\end{enumerate}
\end{Def}

\noindent
If a group $G$ acts on a Riemannian manifold $M$
properly discontinuously by isometries, then
$$
\pi:M\longrightarrow M/G
$$
is an example of a differentiable covering orbifold.

Given point $x$ of an orbifold $Q$,
there is a well-defined
group $G_x$ associated to it.
Let $U = \widetilde{U}/G$ be a local open
coordinate
neighborhood of $x\in  Q$. Then the isotropy subgroup
of $G$ that stabilizes any inverse image of $x$ in $U$ is
unique up to conjugation. We define $G_x$ to be this
isotropy group. When  the isotropy group of $x$ is non-trivial,
then $x$ is said to be a {\em singular} point of the orbifold. The set
of non-singular points is  open and dense in the underlying space $X(Q)$.
An \emph{orbifold cell-decomposition}
of an orbifold is a cell-decomposition of $X(Q)$ such that the group $G_x$ is
the same along each stratum. We denote by $G_C$ the
group associated with a cell $C$.

Thurston extended the notion of the \emph{Euler characteristic} to orbifolds.
\begin{Def}\label{def: orbifold Euler characteristic}
 If  an orbifold $Q$ admits an orbifold
cell-decomposition, then we define the
\emph{Euler characteristic} by
\begin{equation}\label{eq: euler-thurston}
\rchi(Q) = \sum_{C: \text{cell}} (-1)^{\text{dim}(C)}
\frac{1}{|G_C|} .
\end{equation}
\end{Def}
\noindent
The next theorem gives us a useful method to compute
the Euler characteristic.

\begin{thm}\label{thm: Euler characteristic and covering}
Let
$$
\pi: Q_0\longrightarrow Q_1
$$
be a covering orbifold. We define the
\emph{sheet number} of the covering $\pi$
to be the cardinality $k = |\pi^{-1}(y)|$ of the
preimage $\pi^{-1}(y)$ of a non-singular point
$y\in Q_1$. Then
\begin{equation}\label{eq: orbifold covering}
\rchi(Q_1) = \frac{1}{k} \rchi(Q_0) .
\end{equation}
\end{thm}
\begin{proof} We first observe that
for an \emph{arbitrary} point $y$ of $Q_1$, we have
$$
k = \sum_{x: \pi(x) = y}
\frac{|G_x|}{|G_y|} .
$$
Let
$$
Q_1 = \coprod_j C_j
$$
be an orbifold cell-decomposition of $Q_1$, and
$$
\pi^{-1}(C_j) = \coprod_i C_{ij}
$$
a division of the preimage of $C_j$
 into its connected components. Then
\begin{equation*}
\begin{split}
k\rchi(Q_1) &= k \sum_j (-1)^{\text{dim}(C_{j})}
\frac{1}{|G_{C_{j}}|} \\
&= \sum_j (-1)^{\text{dim}(C_{j})}
\sum_i \frac{|G_{C_{j}}|}{|G_{C_{ij}}|} \frac{1}{|G_{C_{j}}|}\\
&= \sum_{ij} (-1)^{\text{dim}(C_{ij})}
\frac{1}{|G_{C_{ij}}|} \\
&= \rchi(Q_0) .
\end{split}
\end{equation*}
\end{proof}

\begin{cor}\label{cor: orbifold Rn mod G}
Let $G$ be a finite subgroup of $\mathfrak{S}_n$
that acts on $\mathbb{R}_+ ^n$ by permutation of
the coordinate axes. Then $\mathbb{R}_+ ^n/G$
is a differentiable orbifold and
\begin{equation}\label{eq: Rn mod G}
\rchi\left(\mathbb{R}_+ ^n/G\right) = \frac{(-1)^n}{|G|} .
\end{equation}
\end{cor}
\begin{rem}
We note that in general
$$
\rchi\left(\mathbb{R}_+ ^n/G\right) \ne \frac{(-1)^n}{|G|},
$$
unless $G$ acts on $\mathbb{R}_+ ^n$ faithfully.
\end{rem}

\begin{ex}\label{ex: Rn mod Sn}
Let us study the quotient space
$\mathbb{R}_+ ^n/\mathfrak{S}_n$. We denote by
$$
\Delta(123\cdots n)
$$
the interior of a regular $n$-hyperhedron of
$(n-1)$ dimensions. Thus
$\Delta(12)$ is a line segment,
$\Delta(123)$ is an equilateral triangle, and
$\Delta(1234)$ is a regular tetrahedron. The space
$\mathbb{R}_+ ^n$ is a cone over $\Delta(123\cdots n)$:
$$
\mathbb{R}_+ ^n = \Delta(123\cdots n)
\times \mathbb{R}_+ .
$$
The closure $\overline{\Delta(123\cdots n)}$
has $n$ vertices $x_1, \cdots, x_n$. Let
$x_{12}$ be the midpoint of the line segment
$\overline{x_1x_2}$, $x_{123}$ the barycenter of
the triangle $\triangle{x_1x_2x_3}$, etc.,
and $x_{123\cdots n}$ the barycenter of
$\Delta(123\cdots n)$.

The $(n-1)$-dimensional region
\begin{equation}\label{RnmodSn}
F =
CH(x_1,x_{12},x_{123},\cdots ,x_{123\cdots n}),
\end{equation}
which is the convex hull of the set of $n$ points
$\{x_1,x_{12},x_{123},\cdots ,x_{123\cdots n}\}$,
is the fundamental domain of the $\mathfrak{S}_n$-action
on $\Delta(123\cdots n)$ induced by permutation of vertices.
It can be considered as a cell complex of
the orbifold $\mathbb{R}_+ ^n/\mathfrak{S}_n$.
It has $\binom{n-1}{k}$ $k$-cells for every
$k$
(Figure~\ref{fig: tetrahedron}).

\begin{figure}[htb]
\centerline{\epsfig{file=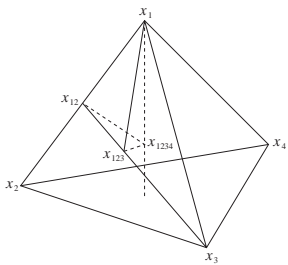, width=1.5in}}
\caption{$\Delta(1234)$.}\label{fig: tetrahedron}
\end{figure}

The
isotropy group of each cell is easily calculated.
For example, the isotropy group of the 2-cell
$CH(x_{12}x_{123}x_{123\cdots n})$
is
$$
\mathfrak{S}(12)\times \mathfrak{S}(456\cdots n),
$$
where $\mathfrak{S}(abc\cdots z)$ is the permutation
group of the specified letters. The definition of the
Euler characteristic (\ref{eq: euler-thurston}) and
a computation using (\ref{eq: orbifold covering}) gives
an interesting combinatorial identity
\begin{equation}\label{eq: Rn mod Sn}
\begin{split}
\rchi\left(\mathbb{R}_+ ^n/\mathfrak{S}_n\right) =
&-\rchi\left(\Delta(123\cdots n)/\mathfrak{S}_n\right)\\
&= -\sum_{k=0} ^{n-1} (-1)^k \sum_{\substack{
m_0 + m_1 + \cdots + m_k = n\\
m_0\ge 1, m_1\ge 1, \cdots, m_k\ge 1}}
\frac{1}{m_0!m_1!\cdots m_k!}\\
&= \frac{(-1)^n}{n!} .
\end{split}
\end{equation}
The $\mathfrak{S}_n$-action of the cell-decomposition
of
$$
\Delta(123\cdots n)/\mathfrak{S}_n
$$
gives a cell-decomposition of
$\Delta(123\cdots n)$ itself, and hence a
cell-decomposition  of $\mathbb{R}_+ ^n$.
We call this cell-decomposition the
\emph{canonical cell-decomposition} of
$\mathbb{R}_+ ^n$, and denote it by
$\square(\mathbb{R}_+ ^n)$. For every subgroup
$G\subset \mathfrak{S}_n$, the fixed point set of
an element of $G$ is one of the cells of
$\square(\mathbb{R}_+ ^n)$. In particular,
$\square(\mathbb{R}_+ ^n)$ induces a cell-decomposition
of the orbifold $\mathbb{R}_+ ^n/G$, which we call
the \emph{canonical orbifold cell-decomposition} of
$\mathbb{R}_+ ^n/G$.
\end{ex}

\section{The Space of Metric Ribbon Graphs}
\label{sect: metric}

The goal of this section is to show that the
space of all metric ribbon graphs with
fixed Euler characteristic and  number of
boundary components forms a
differentiable  orbifold.
The metric ribbon graph
space could
a priori have complicated
singularities, but it  turns out that it
only has quotient singularities given
by certain finite group actions on Euclidean spaces
of a fixed dimension.
This is due to the behavior of the local
deformations of a metric ribbon graph.
The deformations of a
metric ribbon graph which we will discuss below
are related to
certain questions in computer science. We refer to
\cite{STT} for more detail.

Let $RG_{g,n}$ denote the set of all isomorphism classes
of
connected ribbon graphs
$\Gamma $ with no vertices of degree less than $3$
such that
\begin{equation}
\label{eq: topology of Gamma}
\begin{cases}
\rchi(\Gamma ) = v(\Gamma) - e(\Gamma) = 2-2g-n\\
b(\Gamma ) = n,
\end{cases}
\end{equation}
where $v(\Gamma)$, $e(\Gamma)$ and
$b(\Gamma)$ denote the
number of vertices, edges and boundary components
 of $\Gamma$,
respectively.
If an edge $E$ of $\Gamma $
is incident to two {\em distinct} vertices $V_1$ and
$V_2$, then we can construct another
ribbon graph $\Gamma' \in
RG_{g,n}$
by removing the edge $E$ and joining the vertices $V_1$
and $V_2$ to a single vertex, with the cyclic order of the
joined vertex determined by the cyclic order of
the edges incident to $V_1$ starting from the edge
following $E$ up to the edge preceding $E$, followed by the
edges incident to $V_2$ starting with the edge following $E$
and ending with the edge preceding $E$.
The ribbon graph $\Gamma' $ is called a
\emph{contraction} of $\Gamma $.
A \emph{partial ordering} can be
introduced into $RG_{g,n}$
 by defining
\begin{equation}\label{eq: partial ordering of RG}
\Gamma_2  \prec \Gamma_1
\end{equation}
if $\Gamma_2 $ is obtained by a
series of contractions applied to
$\Gamma_1 $.
Since contraction decreases
the number of edges and vertices by one,
 a graph with only
one vertex is a minimal graph, and a {\em trivalent} graph
(a graph with only degree $3$ vertices)   is a maximal
element. Every graph can be
obtained from a trivalent graph by applying
a series of contractions.

\begin{figure}[htb]
\begin{center}
\epsfig{file=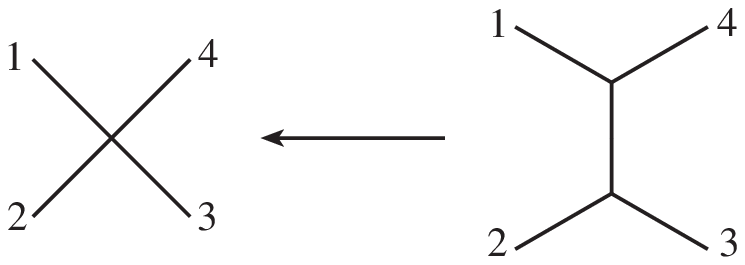, width=1.5in}\\
\epsfig{file=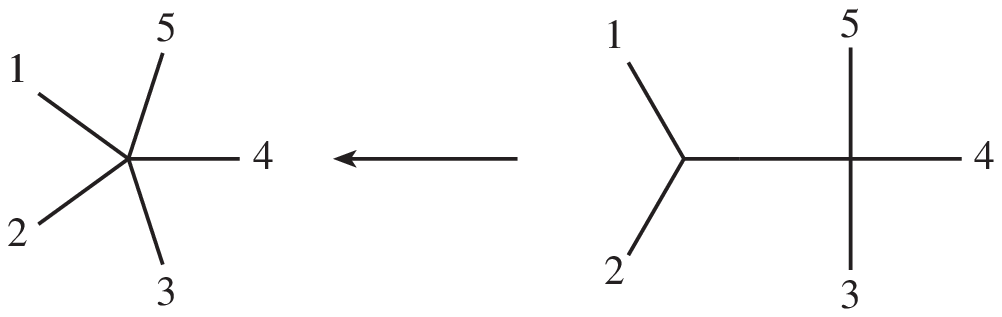, width=1.5in}\\
\epsfig{file=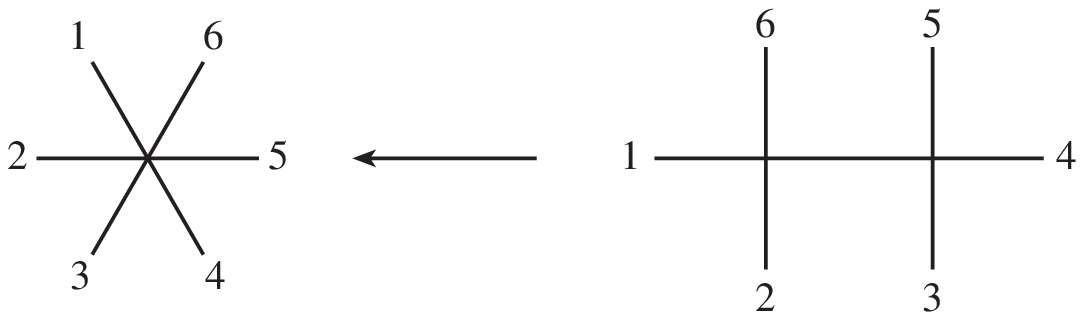, width=1.5in}
\end{center}
\caption{The arrow indicates contraction of
an edge. The inverse direction is expansion of a vertex.}
\label{fig: expansion}
\end{figure}

The inverse operation to the contraction of a ribbon
graph is \emph{expansion}. Every vertex
of degree $d\ge 4$ of a ribbon graph $\Gamma$
can be \emph{expanded} by adding a new edge
as shown in Figure~\ref{fig: expansion}.

\begin{figure}[htb]
\begin{center}
\epsfig{file=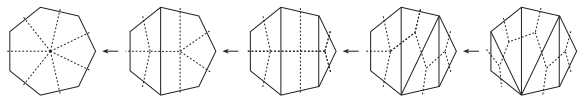, width=3in}
\end{center}
\caption{A series of
expansions of a vertex of degree $7$ and the
dual graphs.}
\label{fig: dual graph}
\end{figure}

In the process of expansion
of an ribbon graph $\Gamma$, we identify two
expanded graphs if there is a ribbon graph isomorphism
from one to the other that preserves all the original
half-edges of $\Gamma$.
 Thus when we expand a vertex of degree
$d\ge 4$,  there are $d(d-3)/2$ ways of expanding
it by adding an edge. The situation is easier to
understand by looking at the \emph{dual graph}
of Figure~\ref{fig: dual graph}, where the arrows
again indicate the contraction map.

Consider the portion of a ribbon graph  consisting of
a vertex of degree $d\ge 4$  and
$d$ half-edges labeled by the numbers $1$ through $d$.
We denote this portion by ${\ast_d}$.
The dual graph of ${\ast_d}$ is a convex polygon with
$d$ sides. The process of expansion by adding an
edge to the center
vertex of ${\ast_d}$ corresponds to drawing a
diagonal line between two
vertices of the $d$-gon, as in
Figure~\ref{fig: dual graph}. The number
$d(d-3)/2$ corresponds to the number of diagonals
in a convex $d$-gon. Expanding the graph
further corresponds to adding another diagonal to
the polygon in such a way that the added diagonal
does not intersect with the existing diagonals except
at the vertices of the polygon.
The expansion process terminates after $d-3$ iterations,
the number non-intersecting
diagonals which can be placed in a convex $d$-gon.
Note that such a maximal expansion is trivalent at the internal vertices,
and its dual defines a triangulation
of the polygon. The number of all triangulations of
a $d$-gon is equal to
$$
\frac{1}{d-1}\binom{2d-4}{d-2},
$$
which is called the \emph{Catalan number}.

A \emph{metric} ribbon graph is a
ribbon graph with a positive
real number assigned to each edge, called the \emph{length} of the edge.
For a ribbon graph $\Gamma\in RG_{g,n}$, the
space of isomorphism
classes of metric ribbon graphs with $\Gamma$ as
the underlying graph is a differentiable orbifold
\begin{equation}
\label{eq: space of metric ribbon graphs}
\frac{\mathbb{R}_+ ^{e(\Gamma)}}{\Aut(\Gamma)} ,
\end{equation}
where the action of $\Aut(\Gamma)$ on $\mathbb{R}_+ ^{
e(\Gamma)}$ is through the natural homomorphism
\begin{equation}
\label{eq: hom of aut to edge permutation}
\phi: \Aut(\Gamma) \longrightarrow
\mathfrak{S}_{e(\Gamma)}.
\end{equation}
For the exceptional graphs
$\Gamma_{\rm{ex}}$ in
Definition~\ref{def: exceptional graphs}, we have
\begin{equation}
\label{eq: exeptional rational cell}
\frac{\mathbb{R}_+ ^{e(\Gamma_{\rm{ex}})}}
{\Aut(\Gamma_{\rm{ex}})}
= \frac{\mathbb{R}_+ ^{e(\Gamma_{\rm{ex}})}}
{\Aut(\Gamma)/(\mathbb{Z}/2\mathbb{Z})}.
\end{equation}

For integers $g$ and $n$ satisfying
\begin{equation}
\label{eq: gn condition}
\begin{cases}
g\ge 0\\
n\ge 1\\
2-2g-n <0,
\end{cases}
\end{equation}
we define the space of isomorphism classes of
metric ribbon graphs satisfying the
topological condition
(\ref{eq: topology of Gamma}) by
\begin{equation}
\label{eq: space of metric ribbon graphs for gn}
RG_{g,n} ^{\text{met}}
= \coprod_{\Gamma\in RG_{g,n}}
\frac{\mathbb{R}_+ ^{e(\Gamma)}}{\Aut(\Gamma)}.
\end{equation}
Each component (\ref{eq: space of metric ribbon graphs})
of (\ref{eq: space of metric ribbon graphs for gn})
is called a \emph{rational cell} of $RG_{g,n} ^{\text{met}}$.
The rational cells are glued together by
the contraction operation in an
obvious way. A rational cell has a natural quotient
topology.

Let us compute the dimension of
 $RG_{g,n} ^{\text{met}}$.
We denote by $v_j(\Gamma)$ the number of
vertices of a ribbon graph $\Gamma$
 of degree $j$. Since
these numbers satisfy
\begin{equation*}
\begin{split}
-(2 - 2g - n) &= -v(\Gamma) + e(\Gamma)\\
&= -\sum_{j\ge 3} v_j(\Gamma) +
\frac{1}{2} \sum_{j\ge 3} j v_j(\Gamma) \\
&= \sum_{j\ge 3} \left(\frac{j}{2}-1\right)
 v_j(\Gamma) ,
\end{split}
\end{equation*}
the number $e(\Gamma)$ of edges takes its maximum
value when all vertices have degree $3$. In that case,
$$
3v(\Gamma) =2 e(\Gamma)
$$
holds, and we have
\begin{equation}\label{eq: ribbon graph dimension}
\dim (RG_{g,n} ^{\text{met}})
= \max_{\Gamma\in RG{g,n}} (e(\Gamma)) = 6g-6+3n .
\end{equation}

To prove that $RG_{r,s} ^{\rm{met}}$ is a
differentiable orbifold,
we need to show that for every
element $\Gamma _{\rm{met}} \in RG_{r,s} ^{\rm{met}}$,
there is an open neighborhood
$U_{\epsilon}(\Gamma _{\rm{met}})$ of
$\Gamma _{\rm{met}}$,
an open disk
$\widetilde{U}_{\epsilon}(\Gamma _{\rm{met}})
\subset \mathbb{R}^{6g-6+3n}$, and a finite group
$G_{\Gamma}$ acting on
$\widetilde{U}_{\epsilon}(\Gamma _{\rm{met}})$
through orthogonal transformations such that
$$
\widetilde{U}_{\epsilon}(\Gamma _{\rm{met}})/
G_{\Gamma}
\cong
U_{\epsilon}(\Gamma _{\rm{met}}).
$$

\begin{Def}\label{def: epsilon neighborhood}
Let
$\Gamma _{\rm{met}} \in RG_{g,n} ^{\rm{met}}$
be a metric ribbon graph, and $\epsilon >0$
a positive number smaller than the half of the length
of the shortest
edge of $\Gamma _{\rm{met}}$.
The \emph{$\epsilon$-neighborhood}
$U_{\epsilon}(\Gamma _{\rm{met}})$ of
$\Gamma _{\rm{met}}$ in
$RG_{g,n} ^{\rm{met}}$ is
the set of all metric ribbon graphs
$\Gamma' _{\rm{met}}$
that satisfy the following conditions:
\begin{enumerate}
\item
$\Gamma \preceq \Gamma'$.
\item The  edges of $\Gamma' _{\rm{met}}$
that are contracted in $\Gamma _{\rm{met}}$
have length less than
$\epsilon$.
\item Let $E'$ be an
 edge of $\Gamma' _{\rm{met}}$
that is not contracted and corresponds to an edge $E$ of
$\Gamma _{\rm{met}}$ of length $L$. Then the length
$L'$ of $E'$ is in the range
$$
L-\epsilon < L' < L+\epsilon .
$$
\end{enumerate}
\end{Def}

\begin{rem} The length of an
 edge of $\Gamma'_{\rm{met}}$
 that is not
contracted in $\Gamma_{\rm{met}}$ is
 greater than $\epsilon$.
\end{rem}

The topology of the space $RG_{g,n} ^{\rm{met}}$ is
defined by these $\epsilon$-neighborhoods. When
$\Gamma_{\rm{met}}$ is trivalent, then
$U_{\epsilon}(\Gamma _{\rm{met}})$ is the
$\epsilon$-neighborhood of
$\mathbb{R}_+ ^{e(\Gamma)}$ in the usual sense.

\begin{Def}
\label{def: space of expansions}
Let $\Gamma\in RG_{g,n}$ be a ribbon graph and
$\Gamma_{\mathcal{E}}$ its edge-refinement.
  We choose a labeling
of all edges of $\Gamma_{\mathcal{E}}$, i.e.,
the half-edges of
$\Gamma$.
The set $X_{\succeq\Gamma}$ consists of
$\Gamma$ itself and all its expansions.
Two expansions are identified if there is
a ribbon graph isomorphism of one expansion
to the other that preserves
the original  half-edges coming from
$\Gamma_{\mathcal{E}}$.
The
\emph{space of metric expansions} of $\Gamma$,
which is denoted by
$X_{\succeq\Gamma} ^{\rm{met}}$, is the set of all
 graphs in $X_{\succeq\Gamma}$
 with a metric on each
edge.
\end{Def}

To understand the structure of
$X_{\succeq\Gamma} ^{\rm{met}}$,
let us consider the expansion process of a vertex of
degree $d\ge 4$. Since expansion is essentially a
local operation, the whole picture can be seen from this
local consideration.
Let ${\ast_d}$ denote the tree graph
consisting of a single vertex of degree $d$
with  $d$ half-edges attached to it.
Although ${\ast_d}$ is not the
type of ribbon graph we are considering, we can define
the space $X_{\succeq{\ast_d}} ^{\rm{met}}$
of metric expansions of ${\ast_d}$ in the same way as
in Definition~\ref{def: space of expansions}.
 Since the edges of
${\ast_d}$ correspond to half-edges of our ribbon graphs,
we do not assign any metric to them.
Thus $\dim(X_{\succeq{\ast_d}} ^{\rm{met}}) = d-3$.
As we have noted in
Figure~\ref{fig: dual graph},
the expansion process of ${\ast_d}$
can be more effectively visualized by looking at the dual polygon.
A maximal expansion corresponds to a triangulation of
the starting $d$-gon by non-intersecting diagonals.
Since there are $d-3$ additional edges in a maximally
expanded
tree graph, each maximal graph is a metric tree homeomorphic to
$\mathbb{R}_+ ^{d-3}$. There is a set of
$d-4$ non-intersecting diagonals in a $d$-gon that
is obtained by removing one diagonal from a triangulation
$T_1$ of the $d$-gon, or removing another diagonal
from another triangulation $T_2$. The transformation
of the tree graph
corresponding to $T_1$ to the tree
corresponding to $T_2$ is the so-called
\emph{fusion move}. If we consider the trivalent
trees as \emph{binary trees}, then the fusion move
is also known as \emph{rotation} \cite{STT}.

\begin{figure}[htb]
\begin{center}
\epsfig{file=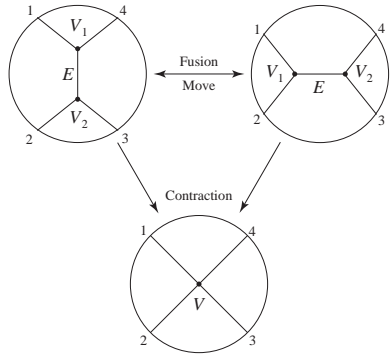, width=2in}
\end{center}
\caption{Fusion move and contraction.}
\label{fig: fusion move}
\end{figure}

Two $(d-3)$-dimensional
cells are glued together along a $(d-4)$-dimensional
cell.
The number of $(d-3)$-dimensional cells in
$X_{\succeq{\ast_d}}$ is equal to the Catalan number.

\begin{thm}
\label{thm: complete expansion scheme}
The space $X_{\succeq{\ast_d}} ^{\rm{met}}$
 is homeomorphic to
$\mathbb{R}^{d-3}$, and its combinatorial structure
defines a cell decomposition of $\mathbb{R}^{d-3}$,
where each cell is a convex cone with vertex at the
origin.  The origin, corresponding to the
graph ${\ast_d}$, is the only $0$-cell of the
cell complex.

The group $\mathbb{Z}/d\mathbb{Z}$
acts on $X_{\succeq{\ast_d}} ^{\rm{met}}$
through orthogonal transformations with respect to
the natural Euclidean structure of
$\mathbb{R}^{d-3}$.
\end{thm}
\begin{rem}
In  \cite{STT},
 the
rotation distance between the top dimensional
cells of $X_{\succeq{\ast_d}} ^{\rm{met}}$
was studied in terms of
hyperbolic geometry, which
has a connection to the structure of binary
search trees.
\end{rem}

\begin{proof}
Draw a convex $d$-gon on the $xy$-plane in
$xyz$-space.
Let $\mathcal{V}$ be the set of vertices of the
$d$-gon, and consider the set
$f\in \mathbb{R}^{\mathcal{V}}$ of all
functions
$$
f:\mathcal{V}\longrightarrow  \mathbb{R} .
$$
An element $f\in \mathbb{R}^{\mathcal{V}}
=\mathbb{R}^d$ can be
represented by its function graph
\begin{equation}
\label{eq: function graph of f}
Graph(f)=\{(V,f(V))\; |\;V\in\mathcal{V}\}
\subset \mathbb{R}^3 .
\end{equation}
Let
us denote by $CH(Graph(f))$ the convex hull of $Graph(f)$
in $\mathbb{R}^3$. If we view the convex hull
from the top, we see a $d$-gon with
a set of non-intersecting
diagonals.

\begin{figure}[htb]
\begin{center}
\epsfig{file=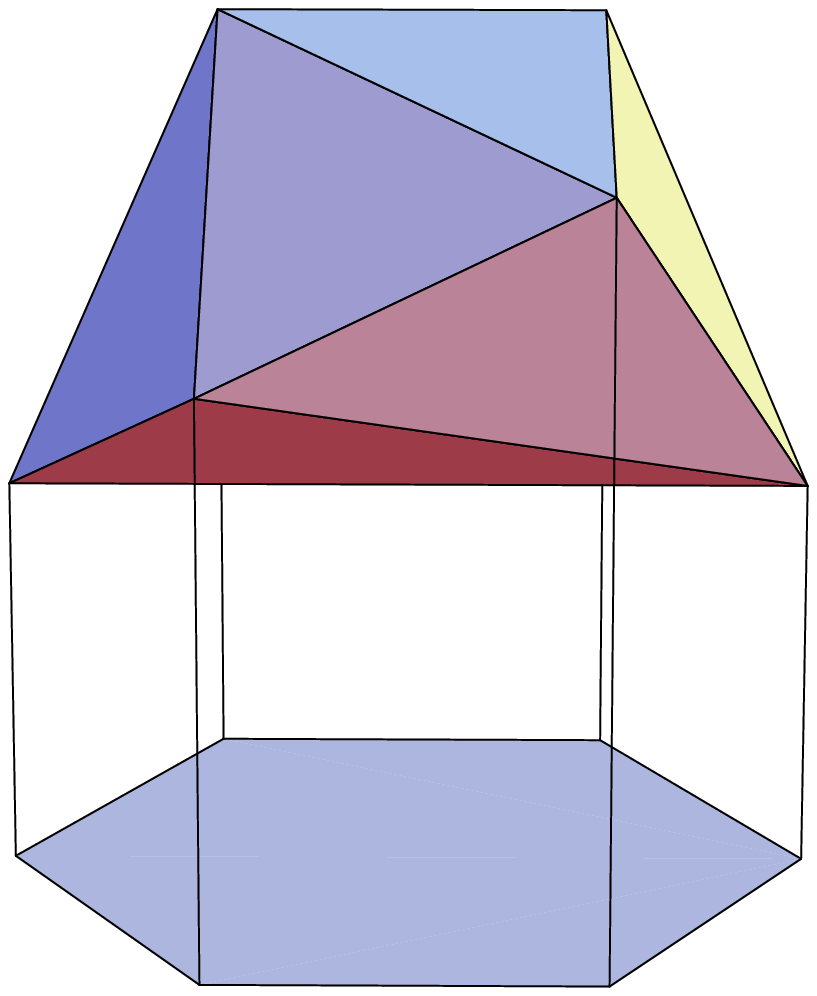, width=1in}\\
\epsfig{file=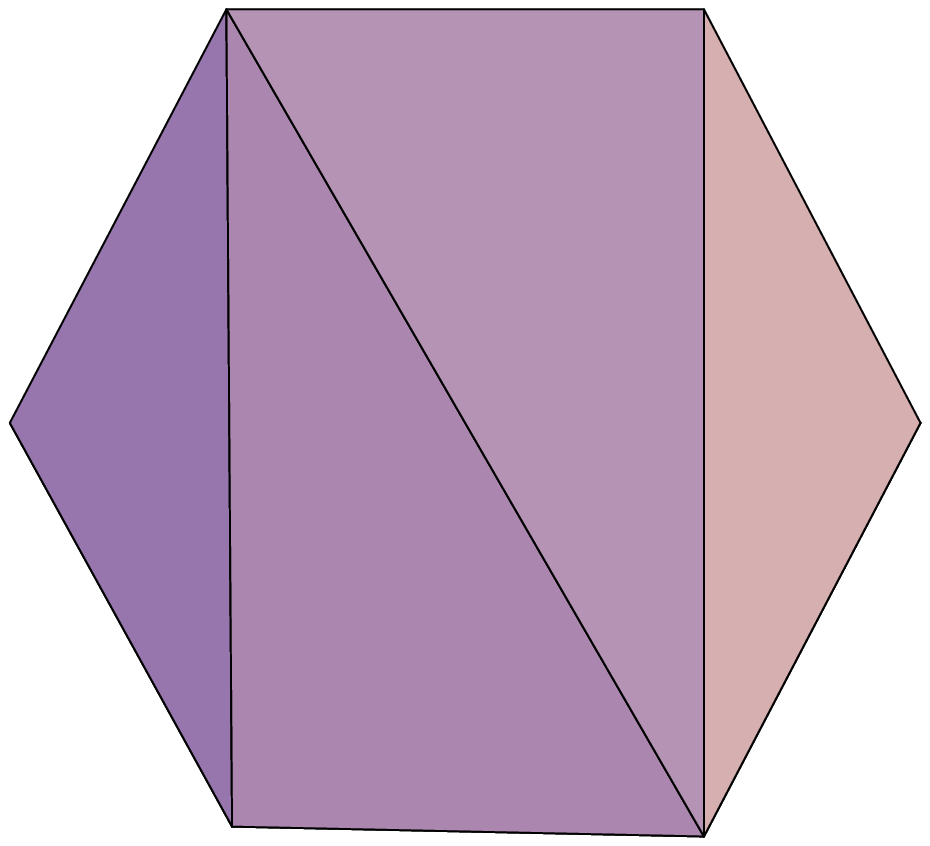, width=1in}
\end{center}
\caption{The convex hull of the function graph of
$f\in \mathbb{R}^{\mathcal{V}}$ and its view
from the top.}
\label{fig: convex hull}
\end{figure}

Viewing the convex hull from the positive direction
of the $z$-axis, we  obtain
a map
\begin{equation}
\label{eq: map xi}
\xi: \mathbb{R}^{\mathcal{V}}
\longrightarrow X_{\succeq{\ast_d}},
\end{equation}
where we identify $X_{\succeq{\ast_d}}$ with the
set of arrangements of non-intersecting
diagonals of a convex $d$-gon.
A generic point of $\mathbb{R}^{\mathcal{V}}$
corresponds to a triangulation of the $d$-gon
as in Figure~\ref{fig: convex hull}, but special
points give fewer diagonals on the
$d$-gon. For example, if $f$ is a constant function,
then the function graph $Graph(f)$ is flat and the top
view of its convex hull is just the $d$-gon
without any diagonals in sight.

This consideration leads us to note that the
map $\xi$ factors through
\begin{equation}
\label{eq: map of Rd-3 to X}
\begin{CD}
\mathbb{R}^{\mathcal{V}}
@>{pr}>>
 \frac{
\mathbb{R}^{\mathcal{V}}}{
\Affine (\mathbb{R}^2, \mathbb{R})
}\\
@|
@VV{\eta}V\\
\mathbb{R}^{\mathcal{V}}
@>{\xi}>>
X_{\succeq{\ast_d}},
\end{CD}
\end{equation}
where
$$
\Affine (\mathbb{R}^2, \mathbb{R})
\isom \mathbb{R}^3
$$
 is the
space of
affine maps of $\mathbb{R}^2$ to $\mathbb{R}$.
Such an affine map
induces a map of $\mathcal{V}$ to $\mathbb{R}$,
but the image is flat and no diagonals are produced
in the $d$-gon.

The map $\eta$ of (\ref{eq: map of Rd-3 to X})
is surjective, because we can explicitly construct
a function $f$ that corresponds to an arbitrary
element of $X_{\succeq{\ast_d}}$.  We also note that
the inverse image
of an $m$-diagonal arrangement ($0\le m\le d-3$)
is a cone of dimension $m$ with vertex at the
origin. It is indeed a convex cone, because
if two points of
\begin{equation}
\label{eq: R d-3}
\frac{\mathbb{R}^{\mathcal{V}}}{
\Affine(\mathbb{R}^2,\mathbb{R})} = \mathbb{R}^{d-3}
\end{equation}
correspond to the same diagonal arrangement of
$X_{\succeq{\ast_d}}$, then every point on the line
segment connecting these two points corresponds
to the same arrangement. To see this, let
$\mathcal{V} = \{V_1,V_2,\cdots,V_d\}$, and let a
function $f\in \mathbb{R}^{\mathcal{V}}$ satisfy
$$
f(V_{d-2}) = f(V_{d-1}) = f(V_d) = 0.
$$
Then $f$ can be thought of an element of the
quotient space (\ref{eq: R d-3}).
Take two such functions
$f$ and $g$ that correspond to the
same $m$-diagonal arrangement of the $d$-gon.
The line segment connecting these two functions
is
\begin{equation}
\label{eq: line segment}
h_t = f + t(g-f),
\end{equation}
where $0\le t\le 1$. This means that the point
$h_t(V_j)\in \mathbb{R}^3$ is on the vertical line
segment connecting $f(V_j)$ and $g(V_j)$ for all
$j=1,2,3,\cdots,d-3.$ Thus the top roof of the
convex hull $CH(Graph(h_t))$ determines the same
arrangement of the diagonals on the $d$-gon as
$CH(Graph(f))$ and $CH(Graph(g))$ do.

Since the inverse image
 of an $m$-diagonal arrangement is an $m$-dimensional
 convex
cone, it is homeomorphic
to $\mathbb{R}_+ ^m$. Hence $X_{\succeq{\ast_d}}$
defines a cell decomposition of $\mathbb{R}^{d-3}$,
which is homeomorphic to
$X_{\succeq{\ast_d}} ^{\rm{met}}$, as claimed.

The convex $d$-gon on the plane can be taken as a
regular $d$-gon centered at the origin.
 The cyclic group
$\mathbb{Z}/d\mathbb{Z}$ naturally acts on
$\mathcal{V}$ through rotations. This action induces
an action on
 $\mathbb{R}^{\mathcal{V}}$  through permutation of axes,
which is an orthogonal transformation
with respect to the standard Euclidean
structure of
 $\mathbb{R}^d$. A rotation of
$\mathcal{V}$ induces a rotation of the horizontal
plane $\mathbb{R}^2$, thus the space
of affine maps
of $\mathbb{R}^2$ to $\mathbb{R}$ is invariant under the
$\mathbb{Z}/d\mathbb{Z}$-action.
The action therefore descends
to the orthogonal complement
$\Affine (\mathbb{R}^2, \mathbb{R})^{\perp}$
in $\mathbb{R}^{\mathcal{V}}$.
Thus $\mathbb{Z}/d\mathbb{Z}$ acts on
$$
X_{\succeq{\ast_d}} ^{\rm{met}}
\isom \Affine (\mathbb{R}^2, \mathbb{R})^{\perp}
\isom \mathbb{R}^{d-3}
$$
by orthogonal transformations with respect to its
natural Euclidean structure.

\end{proof}

\begin{figure}[htb]
\centerline{\epsfig{file=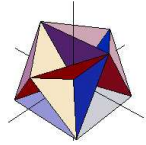, width=1in}
}
\caption{The space of metric expansions
 of a vertex of degree $6$.}
\label{fig: deg6 expansion}
\end{figure}

\begin{ex}
\label{ex: expansion of degree 6}
The space of metric expansions of a vertex of degree
$6$ is a cell decomposition of $\mathbb{R}^3$. There are nine
$1$-cells, twenty-one $2$-cells, and fourteen
$3$-cells. In Figure~\ref{fig: deg6 expansion}, the axes are
depicted in the usual orientation, with the vertical axis representing the $z$
coordinate. The
$\mathbb{Z}/6\mathbb{Z}$-action
on $\mathbb{R}^3$ is generated by
the orthogonal transformation
$$
\begin{pmatrix}
-\frac{1}{2}&0&\frac{\sqrt{3}}{2}\\
0&-1&0\\
-\frac{\sqrt{3}}{2}&0&-\frac{1}{2}
\end{pmatrix}.
$$
\end{ex}

\begin{thm}
\label{thm: space of expansions}
Let $\Gamma\in  RG_{g,n}$. Then
\begin{equation}
\label{eq: space of expansions}
X_{\succeq\Gamma} ^{\rm{met}}
\cong \mathbb{R}_+ ^{e(\Gamma)} \times
\mathbb{R}^{6g-6+3n -e(\Gamma)}.
\end{equation}
The combinatorial structure of
$X_{\succeq\Gamma}$ determines
a cell decomposition of
$\mathbb{R}_+ ^{e(\Gamma)} \times
\mathbb{R}^{6g-6+3n -e(\Gamma)}$.  The group
$\Aut(\Gamma)$ acts on
$X_{\succeq\Gamma} ^{\rm{met}}$
as automorphisms of the cell complex, which are
orthogonal transformations with respect to its
natural Euclidean structure through
the homeomorphism
(\ref{eq: space of expansions}).
The action of $\Aut(\Gamma)$ on
the metric edge space $\mathbb{R}_+ ^{e(\Gamma)}$
may be non-faithful (when $\Gamma$ is exceptional),
but its action on $\mathbb{R}^{6g-6+3n -e(\Gamma)}$
is always faithful except for the case $(g,n) = (1,1)$.
\end{thm}

\begin{proof}
The expansion process of $\Gamma$ takes place
at each vertex of degree $4$ or more. Since we
identify expansions only when there is an isomorphism
fixing all original half-edges coming from $\Gamma$,
the expansion can be done independently at each
vertex. Let $\ast(1),\cdots,\ast(v)$ be the list of
vertices of $\Gamma$ and $d_j$ the degree of
$\ast(j)$. We arrange
the degree sequence of $\Gamma$ as
$$
(\overset{n_3}
{\overbrace{3,3,\cdots, 3}},
\overset{n_4}
{\overbrace{4,4,\cdots, 4}},  \cdots,
\overset{n_m}
{\overbrace{m,m,\cdots, m}}).
$$
Note that
$$
n_3 + n_4 + \cdots + n_m =v =  v(\Gamma)
$$
is the number of vertices of $\Gamma$.
Then
$$
X_{\succeq\Gamma} ^{\rm{met}} =
\mathbb{R}_+ ^{e(\Gamma)} \times
\prod_{j=1} ^v X_{\succeq\ast(j)} ^{\rm{met}},
$$
and the second factor is homeomorphic
to
$$
\prod_{j=1} ^v X_{\succeq\ast(j)} ^{\rm{met}}
\isom
\prod_{\mu=3} ^m \left(\mathbb{R}^{\mu -3}
\right)^{n_{\mu}} =
\mathbb{R}^{\rm{codim}(\Gamma)},
$$
where
$$
{\rm{codim}}(\Gamma) =
6g-6+3n -e(\Gamma) = \sum_{\mu=3} ^m
(\mu-3)n_{\mu}.
$$
The group
$$
G(\Gamma) = \prod_{\mu=3} ^m
\mathfrak{S}_{n_{\mu}} \rtimes
\mathbb{Z}/\mu \mathbb{Z}
$$
acts naturally on
$\prod_{j=1} ^v X_{\succeq\ast(j)} ^{\rm{met}}$
through orthogonal transformations
because each
factor $\mathbb{Z}/\mu\mathbb{Z}$ acts
on $X_{\succeq\ast(j)} ^{\rm{met}}$ through
orthogonal transformations if $d_j = \mu$, and
the symmetric group
$\mathfrak{S}_{n_{\mu}}$ acts on
$\left(\mathbb{R}^{\mu -3}
\right)^{n_{\mu}}$ by permutations of factors, which
are also orthogonal transformations.

Since $\Aut(\Gamma)$ is a subgroup of
$G(\Gamma)$, it acts on
$\prod_{j=1} ^v X_{\succeq\ast(j)} ^{\rm{met}}$
through  orthogonal transformations. It's action
on
$\mathbb{R}_+ ^{e(\Gamma)}$ is by permutations
of axes, thus it is also orthogonal in the
standard embedding of
$\mathbb{R}_+ ^{e(\Gamma)}$ into
$\mathbb{R}^{e(\Gamma)}$.
Therefore, $\Aut(\Gamma)$ acts on
$X_{\succeq\Gamma} ^{\rm{met}}$
through orthogonal transformations
with respect to the natural Euclidean
structure of $X_{\succeq\Gamma} ^{\rm{met}}$.

The action of $\Aut(\Gamma)$
on $\mathbb{R}^{\rm{codim}(\Gamma)}$ is faithful
because all half-edges of $\Gamma$ are labeled
in $X_{\succeq\Gamma} ^{\rm{met}}$,
except for the case  $(g,n) = (1,1)$. There
are only two graphs in $RG_{1,1}$, and both are
exceptional.  Thus the $\Aut(\Gamma)$-action on
$X_{\succeq\Gamma} ^{\rm{met}}$ has a redundant
factor $\mathbb{Z}/2\mathbb{Z}$ for $RG_{1,1}$.
\end{proof}

\begin{thm}
\label{thm: RG is an orbifold}
The space
$$
RG_{g,n} ^{\rm{met}}
= \coprod_{\Gamma\in RG_{g,n}}
\frac{\mathbb{R}_+ ^{e(\Gamma)}}{
\Aut(\Gamma)}
$$
of metric ribbon graphs
is a differentiable  orbifold locally modeled by
\begin{equation}
\label{eq: local model}
\frac{X_{\succeq\Gamma} ^{\rm{met}}}{\Aut(\Gamma)} .
\end{equation}
\end{thm}

\begin{proof}
For every ribbon graph ${\Gamma}\in
RG_{g,n}$, there is a natural map
\begin{equation}
\label{eq: local covering}
\widetilde{\mu}_{\Gamma}:
X_{\succeq\Gamma} ^{\rm{met}}
\longrightarrow
RG_{g,n} ^{\rm{met}},
\end{equation}
assigning to each metric expansion of $\Gamma$
its isomorphism class as a metric ribbon graph.
Since the $\Aut(\Gamma)$-action on
$X_{\succeq\Gamma} ^{\rm{met}}$ induces
ribbon graph isomorphisms, the map
(\ref{eq: local covering}) factors through
the map $\mu_{\Gamma}$ of the quotient space:
\begin{equation}
\label{eq: local isomorphism}
X_{\succeq\Gamma} ^{\rm{met}}
\longrightarrow
\frac{X_{\succeq\Gamma} ^{\rm{met}}}{\Aut(\Gamma)}
\overset{\mu_{\Gamma}}{\longrightarrow}
RG_{g,n} ^{\rm{met}} .
\end{equation}
The inverse image
$
\widetilde{\mu}_{\Gamma} ^{-1}
(U_{\epsilon}(\Gamma_{\rm{met}}))
$
of the
$\epsilon$-neighborhood
$U_{\epsilon}(\Gamma_{\rm{met}})$ is
an open subset of $X_{\succeq\Gamma} ^{\rm{met}}$
that is homeomorphic to a disk. We claim that
\begin{equation}
\label{eq: local homeo}
\mu_{\Gamma}:\frac{
\widetilde{\mu}_{\Gamma} ^{-1}
(U_{\epsilon}(\Gamma_{\rm{met}}))}{\Aut(\Gamma)}
\overset{\sim}{\longrightarrow}
U_{\epsilon}(\Gamma_{\rm{met}})
\end{equation}
is a homeomorphism for every
metric ribbon graph $\Gamma_{\rm{met}}$ if
$\epsilon >0$ is chosen sufficiently
small compared to the shortest edge length of
$\Gamma_{\rm{met}}$.

Take a point $\Gamma_{\rm{met}} ^0\in
U_{\epsilon}(\Gamma_{\rm{met}})$, and let
$$
\Gamma_{\rm{met}} ^j \in
\widetilde{\mu}_{\Gamma} ^{-1}
(\Gamma_{\rm{met}} ^0),
\qquad j=1,2,
$$
be two inverse images. The ribbon graph
isomorphism $\widetilde{\alpha}$ that brings
$\Gamma_{\rm{met}} ^1$ to $\Gamma_{\rm{met}} ^2$
preserves the set $K$ of contracting edges. Since
$\Gamma_{\rm{met}} ^j$ modulo the contracting edges $K$
is the graph $\Gamma_{\rm{met}}$, $\widetilde{\alpha}$
induces an automorphism $\alpha\in \Aut(\Gamma)$.
Thus $\widetilde{\alpha}$ factors into the product of
an automorphism $\alpha$ of $\Gamma$ and a
permutation of $K$.
As an element of $X_{\succeq\Gamma} ^{\rm{met}}$,
a permutation of contracting edges in $K$ fixes the element.
Thus
$\Gamma_{\rm{met}} ^2$ is an $\alpha$-image of
$\Gamma_{\rm{met}} ^1$ in
$X_{\succeq\Gamma} ^{\rm{met}}$.
This shows that (\ref{eq: local homeo}) is a
natural bijection.

Since the topology of the space of metric ribbon
graphs is determined by these $\epsilon$-neighborhoods,
the map $\mu_{\Gamma}$ is continuous. Thus for
a small enough $\epsilon$, we have a homeomorphism
(\ref{eq: local homeo}).

The metric ribbon graph space is covered by
 local coordinate patches
\begin{equation}
\label{eq: coordinate covering}
\bigcup_{\Gamma\in RG_{g,n} ^{\rm{met}}}
\mu_{\Gamma}\left(
\frac{
\widetilde{\mu}_{\Gamma} ^{-1}
(U_{\epsilon}(\Gamma_{\rm{met}}))}{\Aut(\Gamma)}
\right)
= RG_{g,n} ^{\rm{met}},
\end{equation}
where
$$
\frac{
\widetilde{\mu}_{\Gamma} ^{-1}
(U_{\epsilon}(\Gamma_{\rm{met}}))}{\Aut(\Gamma)}
\subset
\frac{X_{\succeq\Gamma} ^{\rm{met}}}
{\Aut(\Gamma)}
$$
is a differentiable orbifold. Let
$$
\Gamma_{\rm{met}} ''
\in
\mu_{\Gamma}\left(
\frac{
\widetilde{\mu}_{\Gamma} ^{-1}
(U_{\epsilon}(\Gamma_{\rm{met}}))}{\Aut(\Gamma)}
\right)
\cap
\mu_{\Gamma '}\left(
\frac{
\widetilde{\mu}_{\Gamma'} ^{-1}
(U_{\epsilon}(\Gamma_{\rm{met}} '))}{\Aut(\Gamma ')}
\right)
$$
be a metric ribbon graph in the intersection of two
coordinate patches. Then
$\Gamma ''\succeq \Gamma$ and
$\Gamma ''\succeq \Gamma '$. There is
a small $\delta$ such that
$$
\mu_{\Gamma''}\left(
\frac{
\widetilde{\mu}_{\Gamma''} ^{-1}
(U_{\delta}(\Gamma_{\rm{met}}''))}{\Aut(\Gamma'')}
\right)
\subset
\mu_{\Gamma}\left(
\frac{
\widetilde{\mu}_{\Gamma} ^{-1}
(U_{\epsilon}(\Gamma_{\rm{met}}))}{\Aut(\Gamma)}
\right)
\cap
\mu_{\Gamma '}\left(
\frac{
\widetilde{\mu}_{\Gamma'} ^{-1}
(U_{\epsilon}(\Gamma_{\rm{met}} '))}{\Aut(\Gamma ')}
\right)  .
$$
If we label the edges of
$\Gamma ''$ that are not contracted in $\Gamma$,
then we have an embedding
$$
X_{\succeq\Gamma ''} ^{\rm{met}}
\subset X_{\succeq\Gamma} ^{\rm{met}}
$$
that induces
$$
\frac{
\widetilde{\mu}_{\Gamma''} ^{-1}
(U_{\delta}(\Gamma_{\rm{met}}''))}{\Aut(\Gamma'')}
\subset
\frac{
\widetilde{\mu}_{\Gamma} ^{-1}
(U_{\epsilon}(\Gamma_{\rm{met}}))}{\Aut(\Gamma)} .
$$
These inclusion maps are injective diffeomorphisms
with respect to the natural differentiable structure
of $X_{\succeq\Gamma} ^{\rm{met}}$.
The same is true for $\Gamma''$ and $\Gamma'$. This
implies that the
local coordinate neighborhoods of
(\ref{eq: coordinate covering})
are patched together by diffeomorphisms.
\end{proof}

\begin{rem} The local map $\mu_{\Gamma}$
of
(\ref{eq: local homeo}) is not a homeomorphism
if $\epsilon$ takes a large value.
In particular,
$$
\frac{X_{\succeq\Gamma} ^{\rm{met}}}
{\Aut(\Gamma)}
$$
does not map injectively to $RG_{g,n} ^{\rm{met}}$
via the natural map $\mu_{\Gamma}$.
\end{rem}

\begin{thm}
\label{thm: Euler characteristic of RG}
The Euler characteristic of $RG_{g,n} ^{\rm{met}}$
as an orbifold is given by
\begin{equation}
\label{eq: Euler characteristic of RG}
\rchi(RG_{g,n} ^{\rm{met}})
= \sum_{\Gamma\in RG_{g,n}} \frac{(-1)^{e(\Gamma)}}
{|\Aut(\Gamma)|}, \qquad (g,n) \ne (1,1).
\end{equation}
For $(g,n) = (1,1)$, we have
\begin{equation}
\label{eq: Euler characteristic of RG1,1}
\rchi(RG_{1,1} ^{\rm{met}}) =
\sum_{\Gamma\in RG_{1,1}} \frac{(-1)^{e(\Gamma)}}
{|\Aut(\Gamma)|/2} = -\frac{1}{3} + \frac{1}{2} =
\frac{1}{6}.
\end{equation}
\end{thm}

\begin{proof}
Since the $\Aut(\Gamma)$-action on
$\mathbb{R}_+ ^{e(\Gamma)}$ is through the
representation
$$
\Aut(\Gamma) \longrightarrow
\mathfrak{S}_{e(\Gamma)},
$$
we have the canonical orbifold cell decomposition
of $\mathbb{R}_+ ^{e(\Gamma)}/\Aut(\Gamma)$ defined in
Example~\ref{ex: Rn mod Sn}.
Gluing all these canonical cell decompositions of
the rational cells of the orbifold $RG_{g,n} ^{\rm{met}}$,
we obtain an orbifold cell decomposition of the
entire space $RG_{g,n} ^{\rm{met}}$.
 To determine
the isotropy subgroups of each orbifold cell, we
need the local model
(\ref{eq: local homeo}). We note that
the $\Aut(\Gamma)$-action on
$\widetilde{\mu}_{\Gamma} ^{-1}
(U_{\epsilon}(\Gamma_{\rm{met}}))$ is faithful
if $(g,n) \ne (1,1)$.
If $\Aut(\Gamma)$ acts on $\mathbb{R}_+ ^{e(\Gamma)}$
faithfully, then the contribution of the rational cell
$\mathbb{R}_+ ^{e(\Gamma)}/\Aut(\Gamma)$ to the
Euler characteristic of $RG_{g,n} ^{\rm{met}}$ is
$$
\frac{(-1)^{e(\Gamma)}}
{|\Aut(\Gamma)|}.
$$
But if $\Gamma$ is exceptional, then the rational
cell
$$
\frac{\mathbb{R}_+ ^{e(\Gamma)}}{\Aut(\Gamma)}
=\frac{\mathbb{R}_+ ^{e(\Gamma)}}{\Aut(\Gamma)/
(\mathbb{Z}/2\mathbb{Z})}
$$
is itself  a singular set of
$X_{\succeq\Gamma} ^{\rm{met}}/\Aut(\Gamma)$.
The contribution of the Euler characteristic of
 $RG_{g,n} ^{\rm{met}}$ from this rational cell is thus
$$
 \frac{(-1)^{e(\Gamma)}}
{2\cdot |\Aut(\Gamma)/
(\mathbb{Z}/2\mathbb{Z})|}
= \frac{(-1)^{e(\Gamma)}}
{|\Aut(\Gamma)|}.
$$
Summing all these, we obtain the formula for
the  Euler characteristic.

The case of $(g,n) = (1,1)$ is still different because
all  graphs in
$RG_{1,1}$ are exceptional. The general formula
(\ref{eq: Euler characteristic of RG}) gives
$-1/6 + 1/4 = 1/12$, but since the factor
$\mathbb{Z}/2\mathbb{Z}$ of $\Aut(\Gamma)$ acts
trivially on $X_{\succeq\Gamma} ^{\rm{met}}$,
the factor $2$ has to be modified.
\end{proof}

The computation of
(\ref{eq: Euler characteristic of RG}) was first done
in \cite{Harer-Zagier} using
a combinatorial argument, and then in
\cite{Penner} and \cite{Mulase1995}
using an asymptotic analysis of
Hermitian matrix integrals. The result is

\begin{equation}\label{eq: computation result}
\sum_{\Gamma\in RG_{g,n}}
\frac{(-1)^{e(\Gamma)}}
{|\Aut(\Gamma)|}
= -  \frac{(2g+n-3)!(2g)(2g-1)}{(2g)!  n!}
\zeta(1-2g)
\end{equation}
for every $g\ge 0$ and $n>0$ subject to
$2-2g-n<0$.

Let $ RGB_{g,n}$ denote the set of
isomorphism classes of  connected ribbon graphs with
labeled boundary components subject to the
topological condition (\ref{eq: topology of Gamma}), and
\begin{equation}
\label{eq: RG met with boundary}
 RGB_{g,n} ^{\text{met}}
= \coprod_{\Gamma\in  RGB_{g,n}}
\frac{\mathbb{R}_+ ^{e(\Gamma)}}{
\Aut_{\partial}(\Gamma)}
\end{equation}
 the space of metric ribbon graphs with labeled
boundary components, where
$$
\Aut_{\partial}(\Gamma)
$$
 is the automorphism
group of a ribbon graph $\Gamma$ preserving the
boundary labeling.
The same argument of the previous section applies
without any alteration to show
 that $ RGB_{g,n} ^{\text{met}}$
is a differentiable orbifold locally modeled by
$$
\frac{X_{\succeq\Gamma} ^{\rm{met}}}{
\Aut_{\partial}(\Gamma)}.
$$
The definition of the space of metric expansions
$X_{\succeq\Gamma} ^{\rm{met}}$ does not refer
to the labeling of the boundary components of
a ribbon graph $\Gamma$, but it requires labeling
of all half-edges of $\Gamma$. As we noted at the end
of Section~\ref{sect: ribbon}, labeling of the half-edges
induces an order of boundary components.
Thus every expansion of $\Gamma$ appearing in
$X_{\succeq\Gamma} ^{\rm{met}}$ has a
 boundary labeling that is consistent with
the boundary labeling of $\Gamma$.

\begin{thm}\label{thm: orbifold covering of RG}
For every genus $g\ge 0$
and $n\ge 1$ subject to
(\ref{eq: gn condition}), the natural \emph{forgetful}
projection
\begin{equation}\label{eq: forget}
pr:  RGB_{g,n} ^{\text{met}}
\longrightarrow
 RG_{g,n} ^{\text{met}}
\end{equation}
is an orbifold covering of degree $n!$.
\end{thm}
\begin{proof}
Let $\Gamma$ be a ribbon graph.
We  label the boundary components
of $\Gamma $, and denote by $B$ the set of all
permutations of the boundary components. The
cardinality $|B|$ of $B$ is $n!$. The automorphism group
$\Aut(\Gamma )$ acts on the set $B$, and by
definition the isotropy subgroup of $\Aut(\Gamma )$
of each element of $B$ is isomorphic to the group
$\Aut_{\partial}(\Gamma)$. The orbit space
$B/\Aut(\Gamma )$ is the set of  ribbon graphs
with labeled boundary. Thus the inverse image
of the local model
$X_{\succeq\Gamma} ^{\rm{met}}/\Aut(\Gamma )$ by
$pr^{-1}$ is the disjoint union of
$|B/\Aut(\Gamma )|$ copies of
$X_{\succeq\Gamma} ^{\rm{met}}/
\Aut_{\partial}(\Gamma)$:
\begin{equation}
pr^{-1} \left(
\frac{X_{\succeq\Gamma} ^{\rm{met}}}{\Aut(\Gamma )}
\right)
= \overset{|B/\Aut(\Gamma )|
{\text{-copies}}}{\overbrace{
\frac{X_{\succeq\Gamma} ^{\rm{met}}}{
\Aut_{\partial}(\Gamma)}
\coprod
\cdots
\coprod
\frac{X_{\succeq\Gamma} ^{\rm{met}}}{
\Aut_{\partial}(\Gamma)}}} .
\end{equation}
Since the projection restricted to each local model
$$
pr: \frac{X_{\succeq\Gamma} ^{\rm{met}}}{
\Aut_{\partial}(\Gamma)}
\longrightarrow
\frac{X_{\succeq\Gamma} ^{\rm{met}}}{\Aut(\Gamma )}
$$
is an orbifold covering of degree
$|\Aut(\Gamma )/\Aut_{\partial}(\Gamma)|$,
the map
\begin{equation}\label{eq: local orbifold covering}
pr_{\Gamma }: pr^{-1} \left(
\frac{X_{\succeq\Gamma} ^{\rm{met}}}{\Aut(\Gamma )}
\right)
\longrightarrow
\frac{X_{\succeq\Gamma} ^{\rm{met}}}{\Aut(\Gamma )}
\end{equation}
is  an orbifold covering of degree
$$
|B/\Aut(\Gamma )|\cdot
|\Aut(\Gamma )/\Aut_{\partial}(\Gamma)|
= |B| = n! .
$$
Since the projection of (\ref{eq: forget}) is just
a collection of $pr_{\Gamma }$ of
(\ref{eq: local orbifold covering}),
$$
pr: RGB_{g,n} ^{\rm{met}}
\longrightarrow {RG}_{g,n} ^{\rm{met}}
$$
is an orbifold covering of degree $n!$ as desired.
\end{proof}

As an immediate consequence, we have
\begin{cor}
\label{cor: Euler characteristic of RG with boundary}
The Euler characteristic of
$ RGB_{g,n} ^{\rm{met}}$
is given by
\begin{equation}
\label{eq: Euler characteristic of RG with boundary}
\rchi( RGB_{g,n} ^{\rm{met}})
= n!\cdot \rchi(RG_{g,n} ^{\rm{met}}).
\end{equation}
\end{cor}

\section{Strebel differentials on Riemann surfaces}
\label{sect: strebel}

A Riemann surface is a patchwork of complex domains. Let us ask
the
question in the opposite direction:
If we are \emph{given} a   compact
Riemann surface, then how can we find
coordinate patches that represent the
complex structure?
In this section we give a canonical
coordinate system
 on a Riemann surface once
 a finite number of points on the surface
and the same number of positive
real numbers are chosen.
 The key technique is the theory of
Strebel differentials \cite{Strebel}. Using Strebel
differentials, we can encode the holomorphic
structure of a Riemann surface in the combinatorial
data of ribbon graphs.

Let $C$ be a compact Riemann surface. We choose a
finite set of labeled points
$\{p_1, p_2, \cdots, p_n\}$ on $C$, and call them
\emph{marked points} on the Riemann surface.
The bridge that connects the complex structure
of a Riemann surface and combinatorial data
 is  the
\emph{Strebel differential} on
the Riemann surface.
Let $K_C$ be the canonical sheaf  of $C$. A
holomorphic \emph{quadratic differential}
defined on $C$
is an element of $H^0(C, K_C ^{\tensor 2})$,
where $K_C ^{\tensor 2}$ denotes the symmetric
tensor product of the canonical sheaf. In a local
coordinate $z$ on $C$, a quadratic differential
$q$ is represented by $q = f(z) \big(dz\big)^2$
with a locally defined holomorphic function $f(z)$.
With respect to a coordinate
change $w=w(z)$, the local expressions
$$
q = f(z)\big(dz\big)^2 = g(w)\big(dw\big)^2
$$
transform by
\begin{equation}\label{eq: qdtransform}
f(z) = g(w(z)) \left(\frac{dw(z)}{dz}\right)^2 .
\end{equation}
A \emph{meromorphic} quadratic
differential on $C$ is a holomorphic quadratic differential
$q$ defined on $C$ except for a finite set
$\{p_1,\cdots, p_n\}$
of points of $C$ such that at each \emph{singularity}
$p_j$ of $q$, there is a local expression
$q=f_j(z)\big(dz\big)^2$ with a meromorphic
function $f_j$ that has a pole at $z=p_j$. If $f_j(z)$ has
a pole of order $r$ at $p_j$, then we say $q$ has a pole of
order $r$ at $z=p_j$.

Let $q = f(z) \big(dz\big)^2$
be a meromorphic quadratic differential defined
on $C$. A real parametric curve
\begin{equation}\label{eq: parametric curve}
\gamma : (a,b) \owns t\longmapsto \gamma(t)=z\in C
\end{equation}
parameterized on an open interval $(a,b)$ of a real
axis is  a \emph{horizontal leaf}
(or in the classical terminology, a horizontal
\emph{trajectory}) of $q$
if
\begin{equation}\label{eq: horizontal leaf}
f(\gamma(t))\left( \frac{d\gamma(t)}{dt}\right)^2 >0
\end{equation}
for every $t\in (a,b)$. If
\begin{equation}\label{eq: vertical leaf}
f(\gamma(t))\left( \frac{d\gamma(t)}{dt}\right)^2 <0
\end{equation}
holds instead, then
the parametric curve $\gamma$ of
(\ref{eq: parametric curve}) is called a \emph{vertical
leaf} of $q$.
The collection of all horizontal or vertical leaves
form a \emph{real codimension
1 foliation} on the Riemann surface $C$
minus the singular points and zeroes of $q$.
There are three important examples
of the foliations for our study.

\begin{figure}[htb]
\centerline{\epsfig{file=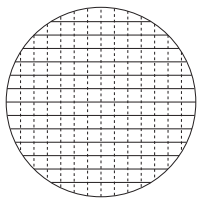, width=1in}}
\caption{Foliations defined by $(dz)^2$.}
\label{fig: hol}
\end{figure}

\begin{ex}\label{ex: qdhol}
Let $q=\big(dz\big)^2$. Then the horizontal lines
$$
\alpha (t) = t + ci, \qquad t\in {\mathbb{R}}
$$
are the horizontal leaves of $q$,
and
$$
\beta (t) = it + c, \qquad t\in {\mathbb{R}}
$$
are the vertical leaves for  every $c\in {\mathbb{R}}$.
Each of these defines a
 simple foliation on the complex plane
$\mathbb{C}$. In Figure~\ref{fig: hol},
horizontal leaves are described by straight lines, and
vertical leaves are indicated by  broken lines.
\end{ex}

If a quadratic differential $q = f(z) (dz)^2$ is holomorphic
and non-zero at $z=z_0$, then on a neighborhood of $z_0$
we can introduce a \emph{canonical coordinate}
\begin{equation}\label{eq: canonical}
w(z) = \int_{z_0} ^z \sqrt{f(z)} dz.
\end{equation}
It follows from (\ref{eq: qdtransform})
that in the $w$-coordinate
the quadratic differential is
given by $q = (dw)^2$. Therefore, the
leaves of $q$ near $z_0$ look
exactly as in Figure~\ref{fig: hol} in the canonical
coordinate.
This explains the classical terminology of horizontal
and vertical trajectories. We remark here that although the
coordinate $w(z)$
is called \emph{canonical}, still there is  an ambiguity
of coordinate change
\begin{equation}
\label{eq: ambiguity of canonical coodinate}
w(z) \longmapsto -w(z) + a
\end{equation}
with an arbitrary complex constant $a$.

Using the canonical coordinate, it is obvious to see the
following:
\begin{prop}
\label{prop: existence of leaves}
Let $S$ be an open Riemann surface and
$q$ a holomorphic quadratic differential on $S$.
Then for every point $p\in S$, there is a unique
horizontal leaf and a vertical leaf passing through
$p$. Moreover, these leaves intersect at a
right angle with respect to the conformal structure
of $S$ near $p$.
\end{prop}

When a holomorphic quadratic differential
has a zero, then the foliation behaves differently.

\begin{figure}[htb]
\centerline{\epsfig{file=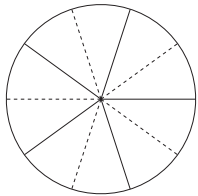, width=1in}}
\caption{Horizontal and vertical
 leaves of $z^{3} (dz)^2$.}\label{fig: zero}
\end{figure}

\begin{ex}\label{ex: qdzero}
Let $q = z^m (dz)^2$. Then $(m+2)$ half rays
$$
\alpha_k : (0,\infty) \owns t \longmapsto
t \cdot\exp\left(\frac{2\pi i k}{m+2}\right) \in \mathbb{C},
\qquad k=0, 1, \cdots, m+1
$$
give the horizontal leaves that have $z=0$
on the boundary
(the straight lines of Figure~\ref{fig: zero}),
and another set of
$(m+2)$ half rays
$$
\beta_k : (0,\infty) \owns t \longmapsto
t \cdot\exp\left( \frac{\pi i +2\pi i k}{m+2}\right)
\in \mathbb{C},
\qquad k=0, 1, \cdots, m+1
$$
gives the vertical leaves
(the broken lines of Figure~\ref{fig: zero}).
\end{ex}

The foliation becomes quite
wild at singularities of $q$. However, the situation is milder
around a quadratic pole with
a negative real coefficient .

\begin{figure}[htb]
\centerline{\epsfig{file=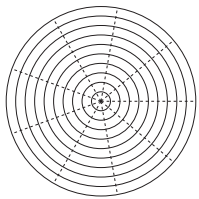, width=1in}}
\caption{Horizontal compact
leaves and vertical leaves
of $-\frac{1}{z^2} (dz)^2$.}\label{fig: pole}
\end{figure}

\begin{ex}\label{ex: qdpole}
Let $q = -\left(\frac{dz}{z}\right)^2$. Then every
concentric circle centered at $0$,
$$
\alpha(t) = r e^{it}, \qquad t\in \mathbb{R}, r>0,
$$
is a horizontal
leaf, and all the half-rays
$$
\beta(t) = t e^{i\theta}, \qquad t > 0,
0\le \theta < 2\pi,
$$
give the vertical leaves. We note that all horizontal
leaves are  compact curves (Figure~\ref{fig: pole}).
\end{ex}

The fundamental theorem we need is:

\begin{thm}[Strebel \cite{Strebel}]
\label{thm: strebel theorem}
Let $g$ and $n$ be integers satisfying that
\begin{equation}\label{eq: gn condition 2}
\begin{cases}
g \ge 0\\
n \ge1\\
2-2g-n <0,
\end{cases}
\end{equation}
and $(C,(p_1,\cdots,p_n))$  a smooth Riemann surface
of genus $g$ with $n$  marked points
$p_1,\cdots,p_n$.
Choose an ordered $n$-tuple
$(a_1, \cdots, a_n)\in \mathbb{R}_+ ^n$ of positive
real numbers. Then there is a unique meromorphic
quadratic differential $q$ on $C$ satisfying the following
conditions:
\begin{enumerate}
\item $q$ is holomorphic on $C\setminus
\{p_1,\cdots,p_n\}$.
\item $q$ has a double pole at each $p_j$, $j=1, \cdots, n$.
\item The union of all noncompact horizontal
leaves forms a closed  subset of $C$ of
measure zero.
\item Every compact horizontal leaf
 $\alpha$ is a simple loop circling around one of
the poles, say $p_j$, and it satisfies
\begin{equation}\label{eq: compactlength}
a_j = \oint_{\alpha} \sqrt{q},
\end{equation}
where the branch of the square root
is chosen so that the integral has
a positive value with respect to the positive
orientation of $\alpha$ that is determined by the
complex structure of $C$.
\end{enumerate}
\end{thm}
 \noindent
This unique quadratic differential is called the
\emph{Strebel differential}.  Note that the
integral (\ref{eq: compactlength}) is automatically
a real number because of (\ref{eq: horizontal leaf}).
Every noncompact horizontal
 leaf of a Strebel differential
defined on $C$ is bounded by zeros of $q$, and
 every zero of degree $m$ of $q$ bounds
 $m+2$ horizontal leaves, as we have
seen in Example~\ref{ex: qdzero}.

Let $\gamma(t)$ be a noncompact horizontal leaf
bounded by two zeros $z_0 = \gamma(t_0)$ and
$z_1 = \gamma(t_1)$ of $q = f(z)(dz)^2$. Then
we can assign a positive real number
\begin{equation}\label{eq: leaf length}
L(\gamma) = \int_{z_0} ^{z_1} \sqrt{q} =
 \int_{t_0} ^{t_1}
\sqrt{f(\gamma(t))} \frac{d\gamma(t)}{dt} dt
\end{equation}
 by choosing a branch of $\sqrt{f(z)}$
near $z_0$ and $z_1$ so that the
integral becomes positive. As before, the integral
is  a real number because
$\gamma$ is a horizontal leaf. We call $L(\gamma)$ the
\emph{length} of the \emph{edge} $\gamma$
with respect to $q$. Note that the length
(\ref{eq: leaf length}) is independent of the choice
of the parameter $t$. The length
is also defined for any compact horizontal leaf by
(\ref{eq: compactlength}). Thus every horizontal leaf has a
uniquely defined length, and hence the Strebel differential
$q$ defines a \emph{measured} foliation on
the open subset of the Riemann surface that
is the complement of the set of zeroes and
poles of $q$.

\begin{figure}[htb]
\centerline{\epsfig{file=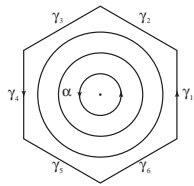, width=1in}}
\caption{A foliated disk with compact horizontal
leaves.}\label{fig: polygon}
\end{figure}

Around every marked point $p_j$ there is a
foliated disk of compact horizontal leaves
with length equal to the prescribed value $a_j$.
As the loop becomes larger in size (but not
in length, because it is a constant), it hits zeroes
of $q$ and the shape becomes a polygon
(Figure~\ref{fig: polygon}).

Let the polygon be an $m$-gon, $\gamma_1, \cdots,
\gamma_m$ the noncompact horizontal leaves
surrounding $p_j$, and $\alpha$ a compact horizontal
 leaf around the point. Then we have
\begin{equation}\label{eq: polygon}
a_j = L(\alpha) = L(\gamma_1) + \cdots L(\gamma_m) .
\end{equation}
We note that some of the $\gamma_j$'s may be the same
noncompact
horizontal leaf
 on the Riemann surface $C$. The collection of
all compact horizontal leaves surrounding $p_j$
forms a punctured disk with its center at $p_j$.
Glue all these punctured disks to noncompact
horizontal leaves and the zeroes of the Strebel
differentials,
and fill  the punctures with points $\{p_1, \cdots, p_n\}$.
Then we obtain a compact surface,
which is the underlying topological surface of the
Riemann surface $C$.

\begin{cor}\label{cor: cor to Strebel}
Let $g$ and $n$ be integers satisfying
(\ref{eq: gn condition 2}),
and
\begin{equation}
\label{eq: pointed Riemann surface with ordered numbers}
(C, (p_1, p_2, \cdots, p_n), (a_1, a_2, \cdots, a_n))
\end{equation}
a nonsingular Riemann surface of genus $g$ with
$n$ marked points and an ordered $n$-tuple
of positive real numbers.
Then there is a unique
cell-decomposition $\square_q$ of $C$
consisting of $v$ $0$-cells, $e$ $1$-cells, and
$n$ $2$-cells, where $v$ is the number of zeroes of
the Strebel differential $q$
associated with
(\ref{eq: pointed Riemann surface with ordered numbers}),
and
\begin{equation*}
e = v-2+2g+n.
\end{equation*}
\end{cor}

\begin{proof}
The $0$-cells, or the \emph{vertices},
 of $\square_q$ are the zeroes of the
Strebel differential $q$ of
(\ref{eq: pointed Riemann surface with ordered numbers}).
The $1$-cells, or the \emph{edges},
 are the noncompact horizontal leaves
that connect the $0$-cells. Since each $1$-cell has a
finite
positive length and the union of all $1$-cells
is closed and has measure
zero on $C$, the number of $1$-cells is finite. The
 union of all
compact horizontal leaves that are homotopic
to $p_j$ (together with the center
$p_j$) forms a $2$-cell, or a \emph{face},
 that is homeomorphic to
a $2$-disk. There are $n$ such $2$-cells.  The
formula for the Euler characteristic
\begin{equation*}
v-e+n = 2-2g
\end{equation*}
determines the number of edges.
\end{proof}

The $1$-skeleton, or
  the union of the $0$-cells and $1$-cells, of the
cell-decomposition $\square_q$ that is defined by
the Strebel differential is a ribbon graph. The
cyclic order of  half-edges at each vertex
is determined by the orientation of the Riemann surface.
A vertex of the graph
that comes from a zero of degree $m$ of the Strebel
differential has degree
$m+2$. Thus the graph we are considering here
does not have any vertices of degree less than
$3$. Since each edge of the graph has the unique
length by (\ref{eq: leaf length}), the graph is a
metric ribbon graph.

Let us give some explicit examples of
the Strebel differentials. We start with a
meromorphic quadratic differential
which is not a Strebel differential, but
nonetheless an important example
because it plays the role of a building
block.

\begin{ex}
\label{ex: meromorphic diff on P1}
Consider the meromorphic quadratic
differential
\begin{equation}
\label{eq: building block}
q_0 = \frac{1}{4\pi^2}\frac{(d\zeta)^2}{\zeta(1-\zeta)}
= \frac{1}{4\pi^2}\left(
\frac{1}{\zeta} + \frac{1}{1-\zeta}
\right)
(d\zeta)^2
\end{equation}
on $\mathbb{P}^1$.
It has simple poles at $0$ and $1$, and
a double pole at $\infty$.
The line segment $[0,1]$ is a horizontal leaf
of length $1/2$.
The whole $\mathbb{P}^1$ minus $[0,1]$ and
infinity is covered with a collection of
compact horizontal leaves which are
confocal ellipses
\begin{equation}
\label{eq: ellipses}
\zeta = a \cos\theta + \frac{1}{2}
+ ib\sin\theta,
\end{equation}
where $a$ and $b$ are positive constants
that satisfy
$$
a^2 = b^2 + \frac{1}{4}.
$$
We have
$$
\frac{1}{4\pi^2}\frac{(d\zeta)^2}{\zeta(1-\zeta)}
= \frac{1}{4\pi^2}(d\theta)^2
$$
under (\ref{eq: ellipses}).
The length of each compact leaf is $1$.
\end{ex}

\begin{figure}[htb]
\centerline{\epsfig{file=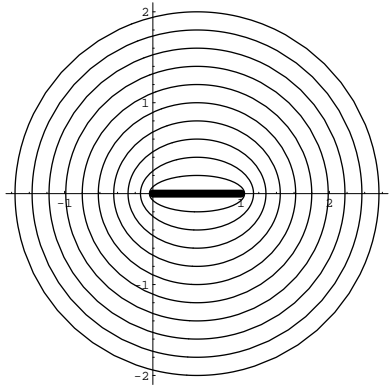, width=2in}}
\caption{Horizontal leaves of
$\frac{1}{4\pi^2}\frac{(d\zeta)^2}{\zeta(1-\zeta)}$.}
\label{fig: ellipses}
\end{figure}

\begin{ex}\label{ex: example m11}
Recall the \emph{Weierstrass elliptic
function}
\begin{equation}\label{eq: p-function}
\begin{split}
\wp(z) &=
\frac{1}{z^2} + \sum_{\substack{(m,n)\in\mathbb{Z} ^2\\
(m,n) \ne (0,0)}}
\left(\frac{1}{(z-m - n\tau)^2}
- \frac{1}{(m + n\tau)^2}\right)\\
&= \frac{1}{z^2} + \frac{g_2 z^2}{20}
+ \frac{g_3 z^4}{28} + \frac{g_2 ^2 z^6}{1200}
+ \frac{3g_2 g_3 z^8}{6160} + \cdots
\end{split}
\end{equation}
defined on the elliptic curve
\begin{equation}
\label{eq: elliptic curve as a quotient}
E_{\tau} = \frac{\mathbb{C}}{\mathbb{Z}
\oplus \mathbb{Z}\tau}
\end{equation}
of modulus $\tau$ with $Im(\tau)>1$,
 where $g_2$ and $g_3$ are defined by
\begin{equation}\label{eq: g2g3}
g_2 =
\sum_{\substack{(m,n)\in\mathbb{Z} ^2\\
(m,n) \ne (0,0)}}
\frac{60}{(m + n\tau)^4},
\quad\text{and}\quad
g_3 =
\sum_{\substack{(m,n)\in\mathbb{Z} ^2\\
(m,n) \ne (0,0)}}
\frac{140}{(m + n\tau)^6} .
\end{equation}
Let $z$ be a coordinate on $E_{\tau}$.
It is customary to write
$$
\omega_1 = 1/2, \quad
\omega_2 = (1 + \tau)/2,\quad
\quad
\omega_3 = \tau/2
$$
and
\begin{equation*}
e_j = \wp(\omega_j), \qquad j = 1, 2, 3.
\end{equation*}
The quantities   $g_2$, $g_3$ and
$e_j$'s satisfy the following relation:
$$
4x^3 - g_2 x - g_3 = 4(x-e_1)(x-e_2)(x-e_3).
$$
Let us consider  the case when $\tau = i =
\sqrt{-1}$.   We have
 $g_2 = 4$, $g_3 = 0$,
$e_1 = - e_3 = 1$ and $e_2  = 0$. In particular,
the elliptic curve is defined over $\mathbb{Q}$.
The Weierstrass $\wp$-function
maps the interior of the
square spanned by $0,\omega_1,\omega_2,
\omega_3$ biholomorphically onto the upper half
plane, and
the boundary of the
square to the real axis
(see for example, \cite{Sansone-Gerretsen}).
A Strebel differential is  given by
\begin{equation}
\label{eq: strebel on Ei}
q = -\frac{4}{\pi^2} \wp(z) (dz)^2 .
\end{equation}
The series expansion of (\ref{eq: p-function})
tells us  that the horizontal leaves near
$0$ are closed loops that are centered at the origin.
The differential $q$ has a double zero at $\omega_2$,
which we see from the Weierstrass differential
equation
$$
\wp'(z) ^2 =
4\wp(z)^3 - 4 \wp(z)  =
 4\left(\wp(z)-e_1\right)
\left(\wp(z)-e_2\right)
\left(\wp(z)-e_3\right).
$$
The real curve $\omega_1 + it$ is a horizontal
leaf, because on the edge
$\overline{\omega_1\omega_2}$ the Weierstrass function
$\wp(z)$ takes values in $[e_2,e_1]$.
The curve $\omega_3  + t$ is also a horizontal
leaf, because on the edge
$\overline{\omega_3\omega_2}$ the function
$\wp(z)$ takes values in $[e_3,e_2]$.
In the above consideration we
used the fact that $\wp(z)$ is an even function:
$$
\wp(z) = \wp(-z).
$$
An extra $\mathbb{Z}/2\mathbb{Z}$ symmetry
comes from the transformation property
$$
\wp(iz) = -\wp(z) .
$$

To prove that (\ref{eq: strebel on Ei}) is indeed
a Strebel differential, we just note that
$q$ is the pull-back of the building
block $q_0$ of (\ref{eq: building block})
via a holomorphic map
\begin{equation}
\label{eq: map of Ei to P1}
\phi: E_i \overset{\wp^2}{
\longrightarrow}\mathbb{P}^1.
\end{equation}
Indeed,
$$
q = -\frac{4}{\pi^2} \wp(z) (dz)^2
= \frac{1}{4\pi^2}\frac{(d\wp^2)^2}
{\wp^2(1-\wp^2)}.
$$
The inverse image of the interval $[0,1]$ is the
degree $4$ ribbon graph with one vertex, two edges and
one boundary component (Figure~\ref{fig: square}).

\begin{figure}[htb]
\centerline{\epsfig{file=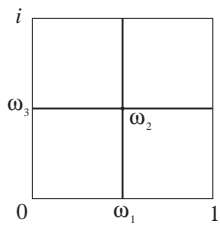, width=1.2in}}
\caption{Elliptic curve of modulus
 $\tau = i$ and a ribbon graph on it.}\label{fig: square}
\end{figure}

Another interesting case
is $\tau = e^{\pi i/3}$, which corresponds to
$g_2 = 0$, $g_3 = 4$, and
\begin{equation*}
e_1 = 1, \quad
e_2 = e^{2\pi i/3},\quad
e_3 = e^{4\pi i/3}.
\end{equation*}
Again the elliptic curve is defined over $\mathbb{Q}$.
The zeroes of $\wp(z)$ are
$$
p=\frac{\omega_1 + \omega_2 + \omega_3}{3}
\quad \text{and}
\quad 2p=\frac{2(\omega_1 + \omega_2 + \omega_3)}{3} .
$$
The Weierstrass function $\wp(z)$ maps the line segment
$\overline{p \omega_1}$ onto $[0,1]$,
 $\overline{p\omega_2}$ to
$\overline{0 e_2}$, and $\overline{p\omega_3}$
to $\overline{0 e_3}$, respectively
(Figure~\ref{fig: triangle}).

\begin{figure}[htb]
\centerline{\epsfig{file=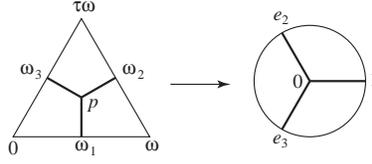, width=2in}}
\caption{Weierstrass $\wp$-function for
$g_2 = 0, g_3 = 4$.}\label{fig: triangle}
\end{figure}

A Strebel differential is given by
\begin{equation}
\label{eq: strebel on Eomega}
q = -\frac{9}{\pi^2}\wp(z) (dz)^2,
\end{equation}
and the non-compact  leaves form
a regular hexagonal network, which possesses
a $\mathbb{Z}/3\mathbb{Z}$-symmetry
(Figure~\ref{fig: hexagon}).

\begin{figure}[htb]
\centerline{\epsfig{file=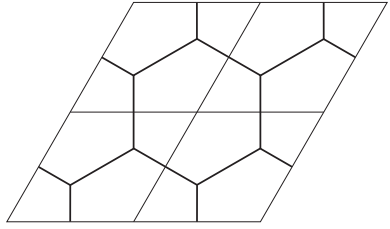, width=2in}}
\caption{A honeycomb.}\label{fig: hexagon}
\end{figure}

We note that (\ref{eq: strebel on Eomega}) is
the pull-back of the building block $q_0$ via
a map
$$
E_{e^{\pi i/3}} \overset{\wp^3}{\longrightarrow}
\mathbb{P}^1.
$$
Indeed, we have
$$
q = -\frac{9}{\pi^2}\wp(z) (dz)^2
= \frac{1}{4\pi^2}\frac{(d\wp^3)^2}
{\wp^3(1-\wp^3)}.
$$
\end{ex}

We can now construct the canonical
coordinate system by
using the Strebel differential once
we give $n$ marked points on a Riemann
surface and an $n$-tuple of real numbers.
Let
$
(C, (p_1, p_2, \cdots, p_n), (a_1, a_2, \cdots, a_n))
$
be the set of data of
(\ref{eq: pointed Riemann surface with ordered numbers}),
and $q$ the Strebel differential
associated with the data. We recall that for every
point of a vertical leaf there is a horizontal leaf
intersecting perpendicularly at the point
(Proposition~\ref{prop: existence of leaves}).
Since the set of compact horizontal leaves
of $q$ forms an open dense subset of $C$, which is
indeed the disjoint union of punctured
open $2$-disks, every
vertical leaf of $q$ extends to one of the points $p_j$.
In particular, a vertical leaf starting at a
zero of $q$ should end at one of the poles.

\begin{thm}
\label{thm: canonical triangulation of a Riemann surface}
The set of all vertical leaves that connect
zeroes and poles of $q$, together with
the cell-decomposition $\square_q$
of $C$ by the noncompact horizontal leaves,
defines a \emph{canonical
triangulation} $\Delta_q$ of $C$.
\end{thm}

\begin{figure}[htb]
\begin{center}
\epsfig{file=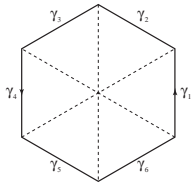, width=1in}
\end{center}
\caption{Triangulation of a polygon}
\label{fig: polygontriang}
\end{figure}

\begin{proof}
The cell-decomposition $\square_q$ of
$C$ defined by the noncompact horizontal leaves
of $q$ defines a polygonalization of $C$. Each polygon
Figure~\ref{fig: polygon} has a unique center, which is
a pole of $q$. The vertical leaves that connect
zeroes and poles supply the edges necessary for
a triangulation of each polygon
(Figure~\ref{fig: polygontriang}).
\end{proof}

\begin{figure}[htb]
\begin{center}
\epsfig{file=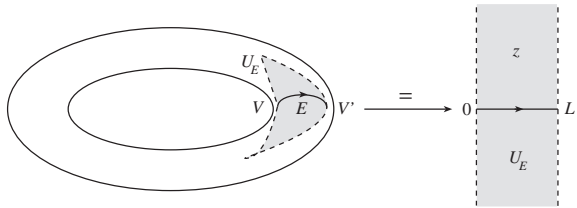, width=3in}
\end{center}
\caption{Triangulation and a canonical coordinate system
of a Riemann surface.}
\label{fig: strip}
\end{figure}

Let $\Gamma_q$ denote a  ribbon graph consisting of
zeroes of $q$ as vertices and noncompact
horizontal leaves of $q$ as edges.  By the property of the
Strebel differential, we have
\begin{equation}\label{eq: ribbon graph of q}
\begin{cases}
\rchi(\Gamma_q) = v(\Gamma_q) -
e(\Gamma_q) = 2-2g\\
b(\Gamma_q) = n.
\end{cases}
\end{equation}
In particular, the closed surface associated with
the ribbon graph $\Gamma_q$ is the
underlying topological
surface of the Riemann surface $C$.
 For every edge $E$ of $\Gamma_q$,
there are two triangles of $\Delta_q$
that share $E$.
Gluing these two triangles along $E$, we obtain
a diamond shape as in Figure~\ref{fig: strip}.
This is the set of all vertical leaves that intersect
with $E$. Let $V$ and $V'$ be  the endpoints of $E$,
and give a direction to $E$ from $V$ to $V'$. We allow
the case that $E$ has only one endpoint. In that
case, we assign an arbitrary direction to $E$. For
a point $P$  in the triangles, the
canonical coordinate
\begin{equation}
\label{eq: canonical coordinate for triangles}
z =z(P) = \int_{V} ^P \sqrt{q}
\end{equation}
maps the diamond shape to a strip
\begin{equation}
\label{eq: infinite strip}
U_E = \{z\in\mathbb{C}\;|\;
0< Re(z) < L\}
\end{equation}
of infinite height and width $L$
in the complex plane, where $L$ is the length of
$E$. We identify the open set $U_E$ as the
union of two triangles on the Riemann surface $C$
by the canonical coordinate $z$ (Figure~\ref{fig: strip}).
The local expression of $q$ on $U_E$ is of course
\begin{equation}
\label{eq: q in UE}
q = (dz)^2 .
\end{equation}

\begin{figure}[htb]
\begin{center}
\epsfig{file=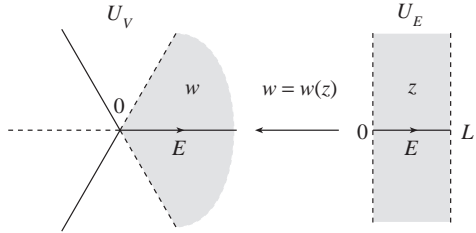, width=2.5in}
\end{center}
\caption{Gluing a strip to a neighborhood of a vertex
by $w = w(z) = z^{2/3}$.}
\label{fig: glue z and w}
\end{figure}

Let the degree of $V$ be $m$. We note that
every quadratic differential has an expression
\begin{equation}\label{eq: q at zero}
q = \frac{m^2}{4} w^{m-2} (dw)^2
\end{equation}
 around a zero of degree $m-2$. So we use
(\ref{eq: q at zero}) as the expression of the
Strebel differential $q$ on an open neighborhood
$U_V$ around $V$ with a coordinate $w$ such that
$V$ is given by $w=0$. On the intersection
$$
U_E \cap U_V ,
$$
we have
\begin{equation}
\label{eq: diff eq z=w}
q = (dz)^2 = \frac{m^2}{4} w^{m-2} (dw)^2
\end{equation}
from (\ref{eq: q in UE}) and (\ref{eq: q at zero}).
Solving this differential equation with the initial
condition that $z=0$ and $w=0$ define the same
point $V$, we obtain the coordinate transform
\begin{equation}\label{eq: z=w}
w = w(z) = c z^{2/m} ,
\end{equation}
where $c$ is an $m$th root of unity.
Thus $U_E$ and $U_V$ are
glued on the Riemann surface $C$ in the way
described in Figure~\ref{fig: glue z and w}.

 Since we have
$$
\oint \sqrt{q} = a_j
$$
around  a quadratic
pole $p_j$, we can choose a local coordinate $u$
on an open disk $U_j$ centered at $p_j$ such that
\begin{equation}\label{eq: q at pole pj}
q = -\frac{a_j ^2}{4\pi^2}\frac{(du)^2}{u^2} .
\end{equation}
The coordinate disk $U_j$, which is the
union of the horizontal
leaves that are zero-homotopic to $p_j$,
can be chosen so that its boundary
 consists of a collection of edges
$E_1, \cdots, E_{\mu}$ for some $\mu$. Let $z_k$
be the canonical coordinate on $U_{E_k}$.
Equations (\ref{eq: q in UE}) and
(\ref{eq: q at pole pj}) give us a differential equation
\begin{equation}
\label{eq: diff eq z=u}
(dz_k)^2 = -\frac{a_j ^2}{4\pi^2}\frac{(du)^2}{u^2} .
\end{equation}
Its solution is given by
\begin{equation}\label{eq: u=z}
u = u(z_k) = \gamma e^{2\pi iz_k/a_j} ,
\end{equation}
where $\gamma$ is a constant of integration.
Since the edges $E_1, \cdots, E_{\mu}$ surround the point
$p_j$, the constant of integration  for each
$z_k$ is arranged so that the solution
$u$ of (\ref{eq: u=z}) covers the entire disk.
The precise form of gluing function of
open sets $U_{E_k}$'s and $U_j$ is given by
\begin{equation}
\label{eq: u=z final}
u = u(z_k) =
\exp \left(2\pi i
\frac{L_1 + L_2 + \cdots + L_{k-1} + z_k}
{L_1 + L_2 + \cdots + L_{\mu}}\right), \qquad
k = 1, 2, \cdots, \mu ,
\end{equation}
where the length $L_k$ satisfies the condition
$$
a_j = L_1 + L_2 + \cdots + L_{\mu} .
$$

\begin{figure}[htb]
\begin{center}
\epsfig{file=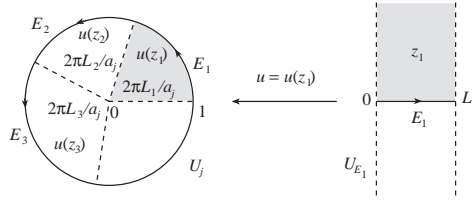, width=2.5in}
\end{center}
\caption{Gluing a strip to a neighborhood of a pole
by $u = u(z_1) = \exp(2\pi i z_1/a_j)$.}
\label{fig: glue u and z}
\end{figure}

The  open coordinate charts  $U_E$'s, $U_V$'s and
$U_j$'s cover the whole Riemann surface $C$.
We call them the \emph{canonical coordinate
charts}.

\begin{Def}
\label{def: canonical coordinate system}
The \emph{canonical coordinate system} of
the data
 $$
(C, (p_1,p_2,\cdots,p_n),(a_1,a_2,\cdots,a_n))
$$
is the covering
\begin{equation}
\label{eq: canonical coordinate covering}
C = \bigcup_{E} U_E \cup \bigcup_{V} U_V\cup
\bigcup_{j=1} ^n U_j
\end{equation}
of the Riemann surface $C$ by the canonical
coordinate charts,
where the union is defined by the gluing functions
(\ref{eq: z=w}) and (\ref{eq: u=z final}).
\end{Def}

We note that the canonical coordinate
$z$ we have chosen for the strip $U_E$ around an
edge $E$ depends on the direction of the edge. If
we use the opposite direction, then the coordinate
changes to
\begin{equation}
\label{eq: change of canonical coordinate}
z \longmapsto L-z,
\end{equation}
where $L$ is the length of $E$, as before. This
change of coordinate does not affect the
differential equations (\ref{eq: diff eq z=w})
and (\ref{eq: diff eq z=u}), because
$$
(dz)^2 = (d(L-z))^2 .
$$

\section{Combinatorial description of the
moduli spaces of Riemann surfaces}
\label{sect: combi}

We have defined the space of
 metric ribbon graphs with labeled boundary components
by
\begin{equation*}
 RGB_{g,n} ^{\rm{met}} =
\coprod_{\Gamma\in RGB_{g,n}}
\frac{\mathbb{R}_{+} ^{e(\Gamma)}}
{\Aut_{\partial}(\Gamma)}.
\end{equation*}
The Strebel theory defines a
map
\begin{equation}\label{eq: Mgn to RGgn}
\sigma : \mathfrak{M}_{g,n}
\times \mathbb{R}_+ ^n
\longrightarrow
 RGB_{g,n} ^{\text{met}} ,
\end{equation}
where $\mathfrak{M}_{g,n}$ is the \emph{moduli
space} of Riemann surfaces of genus $g$ with
$n$ ordered marked points.
In  what follows, we
 prove that
the map $\sigma$ is   bijective.
The case of $(g,n) = (1,1)$ is in many ways
 exceptional. For $(g,n) = (0,3)$, all
Strebel differentials are explicitly computable.
Thus we can construct the identification map
$$
 \mathfrak{M}_{0,3}
\times \mathbb{R}_+^3 =
RGB_{0,3} ^{\text{met}}
$$
directly.  This topic
is studied at the end of this section.
We  also examine the orbifold
covering $RGB_{0,3} ^{\rm{met}}\rightarrow
RG_{0,3} ^{\rm{met}}$ there.

The product group $\mathbb{R}_+ ^n$ acts naturally
on $\mathfrak{M}_{g,n}
\times \mathbb{R}_+ ^n$. Therefore it acts on
the space $RGB_{g,n} ^{\rm{met}}$ through the
bijection $\sigma$. However,
 the  $\mathbb{R}_+ ^n$-action on
$RGB_{g,n} ^{\text{met}}$ is complicated. We give an
example in this section which
shows that the action does not preserve the
rational cells.

\begin{thm}\label{thm: Mgn = RGgn}
There is a natural bijection
\begin{equation*}
\mathfrak{M}_{g,n}
\times \mathbb{R}_+ ^n
=
 RGB_{g,n} ^{\text{met}} .
\end{equation*}
\end{thm}

\begin{proof} The proof breaks down into three steps.
In Step~1, we construct a map
\begin{equation}\label{eq: inversemap}
\coprod_{\Gamma\in RGB_{g,n}}
\mathbb{R}_{+} ^{e(\Gamma)}
\longrightarrow
 \mathfrak{M}_{g,n}\times
\mathbb{R}_+ ^n .
\end{equation}
We then prove that the map descends to
\begin{equation}\label{eq: RGgn to Mgn}
\beta :  RGB_{g,n} ^{\text{met}}
\longrightarrow
\mathfrak{M}_{g,n}\times
\mathbb{R}_+ ^n
\end{equation}
by considering the action of  the
  graph automorphism
groups preserving the boundary
order in Step~2. {F}rom the construction of
Step~1 we will see that $\beta$ is a right-inverse of
the map $\sigma$ of (\ref{eq: Mgn to RGgn}), i.e.,
$\sigma\circ\beta$ is the identity of
$ RGB_{g,n} ^{\text{met}}$.
In Step~3 we prove that $\beta$ is also a left-inverse of
the map $\sigma$.

\begin{step} Our starting point is a  metric ribbon graph
$\Gamma_{\rm{met}}$ with labeled boundary components.
We label all edges, and give an arbitrary
direction to each edge. To each directed edge
$\overrightarrow{E}$ of $\Gamma_{\rm{met}}$,
we assign a strip
$$
U_{\overrightarrow{E}}
 = \{z\in\mathbb{C}\; | \;0< Re(z)
< L\}
$$
of infinite length and width $L$, where $L$ is the
length of $E$
(see Figure~\ref{fig: strip2}).
The open real line segment $(0,L) \subset
U_{\overrightarrow{E}}$
is identified with the edge $\overrightarrow{E}$.
The strip $U_{\overrightarrow{E}}$ has a
complex structure defined by
the
coordinate $z$, and a holomorphic
quadratic differential
$
 (dz)^2
$
 on it. Every horizontal leaf
of the foliation defined by this quadratic differential
 is a  horizontal line of length $L$.
If we use the
opposite direction of $\overrightarrow{E}$, then
$U_{\overrightarrow{E}}$
should be rotated $180^\circ$ about the
real point $L/2$, and the coordinate is changed to
$L-z$.

\begin{figure}[htb]
\begin{center}
\epsfig{file=ribbonfig/figstrip2.eps, width=0.5in}
\end{center}
\caption{A strip of infinite length with horizontal
leaves.}
\label{fig: strip2}
\end{figure}

Let $V$ be  a degree $m$ vertex  of $\Gamma$.
There are $m$ half-edges attached to $V$, although
some of them may belong to the same edge.
Let $1, 2, \cdots, m$ be the cyclic order of
the half-edges chosen at $V$. We give a
direction to each edge by defining the
positive direction to be the one coming out from $V$,
and name the edges
$\overrightarrow{E}_1,
\overrightarrow{E}_2, \cdots,
\overrightarrow{E}_m$. If an edge goes out as a
half-edge number $j$ and comes back as another
half-edge number $k$, then we use the convention that
$\overrightarrow{E}_j = \overleftarrow{E}_k$, where
$\overleftarrow{E}_k$
 denotes the edge $E_k$ with the opposite
direction.
We denote by $L_j$  the length of $E_j$.

Let us place the vertex $V$ at the origin of the $w$-plane.
We glue
a neighborhood of the boundary point $0$ of each of
the strips
$U_{\overrightarrow{E}_1}, \cdots,
U_{\overrightarrow{E}_m}$ together on the $w$-plane
by
\begin{equation}\label{eq: zeroglue}
w = e^{2\pi i (j-1)/m} z_j ^{2/m}, \qquad j=1, 2,\cdots, m.
\end{equation}

An open neighborhood $U_V$ of $w=0$ is covered
by this gluing, if we include the boundary of each
$U_{\overrightarrow{E}_j}$.
It follows from (\ref{eq: zeroglue}) that
the expression of the quadratic differential
$(dz_j)^2$ changes
into
\begin{equation}\label{eq: q-differential at zero}
(dz_j)^2 = \frac{m^2}{4} w^{m-2} (dw)^2
\end{equation}
in the $w$-coordinate
for every $j$. So we define a holomorphic
quadratic differential $q$ on $U_V$
by (\ref{eq: q-differential at zero}).
Note that  $q$
has a zero of degree $m-2$ at $w=0$. At least locally
on $U_V$, the
horizontal leaves of the foliation defined by $q$
that have $V$ as a boundary
point
coincide with the image of the edges $E_1, \cdots,
E_m$ via (\ref{eq: zeroglue}).

Next, let us consider the case when
edges $\overrightarrow{E}_1,
\overrightarrow{E}_2, \cdots,
\overrightarrow{E}_h$
form an oriented boundary component $B$ of
$\Gamma$, where the direction
of $\overrightarrow{E}_k$ is chosen to be
compatible with the orientation of  $B$.
Here again we allow that some of the edges are actually
the same, with the opposite
direction. As before, let $L_k$ be the length of $E_k$,
and put
\begin{equation}\label{eq: looplength}
a_B = L_1 + L_2 + \cdots L_h .
\end{equation}
This time we glue the upper half of the strips
$U_{\overrightarrow{E}_1}, \cdots,
U_{\overrightarrow{E}_h}$
(or the lower half, if the edge has the opposite
direction) into the unit disk of the $u$-plane by
\begin{equation}\label{eq: poleglue}
u = \exp\left(\frac{2\pi i}{a_B} (L_1 + L_2 + \cdots +
L_{k-1} + z_k)\right), \qquad k=1, 2, \cdots, h .
\end{equation}
We note that the entire unit disk on the $u$-plane,
which we denote by $U_B$,
is covered by this gluing, if the boundary
lines of the strips are included.

It follows from this coordinate transform that
\begin{equation}\label{eq: poledifferential}
(dz_k)^2 = -\frac{a_B ^2}{4\pi^2} \frac{(du)^2}{u^2}.
\end{equation}
Thus the holomorphic
quadratic differential $q$ naturally extends to
a meromorphic quadratic differential on the union
$$
U_B\cup \bigcup_{k=1} ^h U_{\overrightarrow{E}_k}
$$
which has a pole of order 2 at $u=0$ with a  negative real
coefficient.
The horizontal leaves of the foliation defined by $q$
 are concentric circles that are centered
 at $u=0$, which correspond to the horizontal
lines on $U_{\overrightarrow{E}_k}$ through
(\ref{eq: poleglue}).
Note that the length of
a compact horizontal leaf around $u=0$ is always $a_{B}$.

Now define
a compact Riemann surface
$C(\Gamma_{\rm{met}})$ by gluing
all the $U_V$'s, $U_B$'s and the strips
$U_{\overrightarrow{E}}$'s by
(\ref{eq: zeroglue}) and (\ref{eq: poleglue}):
\begin{equation}
\label{eq: C Gamma}
C(\Gamma_{\rm{met}})
= \bigcup_{V: \text{ vertex of } \Gamma} U_V
\cup \bigcup_{E: \text{ edge of } \Gamma}
U_{\overrightarrow{E}}
\cup \bigcup_{\substack{B: \text{ boundary}\\
\text{component of } \Gamma}} U_B  .
\end{equation}
Since there are two directions for every edge $E$,
both the upper half part and the lower half part
of a strip $U_{\overrightarrow{E}}$ are included in
the union of all $U_B$'s. Thus the union (\ref{eq: C Gamma})
is compact.
 The Riemann surface $C(\Gamma_{\rm{met}})$ has
$n=b(\Gamma)$   marked points each of which is the
center of the unit disk $U_B$.
The ordering of the boundary components of the
ribbon graph determines an ordering
of the marked points on the Riemann surface.
Attached to
 each marked point we have a positive real number
$a_B$. The Riemann surface
also comes with a meromorphic quadratic
differential whose local expressions are
given by $(dz_j)^2$, (\ref{eq: q-differential at zero}),
and (\ref{eq: poledifferential}).
It is a Strebel differential on
$C(\Gamma_{\rm{met}})$.
The  metric ribbon graph
corresponding to this Strebel differential is, by
construction, exactly the
original graph $\Gamma_{\rm{met}}$, which has
a natural ordering of the boundary components.
Thus we have constructed a map
(\ref{eq: inversemap}).
\end{step}

\begin{step}
Let us  consider the effect of a graph automorphism
$f\in \Aut_{\partial}(\Gamma)$ on (\ref{eq: C Gamma}).
Let $p_1, \cdots, p_n$ be the marked points of
$C(\Gamma_{\rm{met}})$, and
$a_1, \cdots, a_n$ the corresponding positive
numbers. We denote by $E_1, \cdots, E_e$  the edges
of $\Gamma$, and by
$U_{\overrightarrow{E}_1}, \cdots,
U_{\overrightarrow{E}_e}$
the corresponding strips with a choice of direction. Then
the union of the closures of these strips cover the
Riemann surface minus the marked points:
$$
C(\Gamma_{\rm{met}})
\setminus \{p_1,\cdots, p_n\}
= \bigcup_{j=1} ^e
\overline{U}_{\overrightarrow{E}_j} .
$$
A graph automorphism $f: \Gamma_{\rm{met}}
\rightarrow \Gamma_{\rm{met}}$ induces a permutation
of edges and flip of directions,
 and hence a permutation of strips
$U_{\overrightarrow{E}_1}, \cdots,
U_{\overrightarrow{E}_e}$ and a change of
coordinate $z_j$ to $L_j - z_j$. If $f$ fixes a vertex
$V$ of degree $m$,
then it acts on $U_V$ by rotation of angle an
integer multiple of $2\pi/m$, which is a holomorphic
automorphism of $U_V$. The permutation of
vertices induced by $f$ is a holomorphic transformation
of the union of $U_V$'s. Since $f$ preserves the
boundary components of $\Gamma$, it does not
permute $U_B$'s, but it may rotate each $U_B$
following the effect of the permutation of edges.
In this case, the origin of $U_B$, which is one of the
marked points, is fixed, and the orientation of the
boundary is also fixed.
Thus the graph
automorphism induces a holomorphic automorphism of
$C(\Gamma_{\rm{met}})
\setminus \{p_1,\cdots, p_n\}$. This
holomorphic automorphism
preserves the ordering of the marked points.
Thus we conclude that (\ref{eq: inversemap}) descends to
a map $\beta$ which satisfies  $\sigma\circ\beta
= id$.
\end{step}

\begin{step} We still need to show that
$\beta\circ\sigma$ is the identity map of
$\mathfrak{M}_{g,n}\times
\mathbb{R}_+ ^n$, but
this is exactly what we have shown in
the previous section.
\end{step}

 This completes the
proof of Theorem~\ref{thm: Mgn = RGgn}.
\end{proof}

\begin{ex}
Let us consider  the complex
projective line $\mathbb{P} ^1$ with three
ordered marked points, to illustrate the equality
$$
\mathfrak{M}_{0,3} \times \mathbb{R}_{+} ^3
=  RGB_{0,3} ^{\text{met}}
$$
and the covering map
$$
 RGB_{0,3} ^{\text{met}}
\longrightarrow
 RG_{0,3} ^{\text{met}} .
$$
The holomorphic automorphism
group of $\mathbb{P} ^1$ is $PSL(2,\mathbb{C})$, which
acts on $\mathbb{P} ^1$ triply transitively. Therefore,
we have a biholomorphic equivalence
$$
(\mathbb{P} ^1, (p_1,p_2,p_3))
\cong
(\mathbb{P} ^1, (0, 1, \infty)).
$$
 In
other words, $\mathfrak{M}_{0,3}$ is just a point.
Choose a triple $(a_0,a_1,a_{\infty})$ of positive real
numbers. The unique Strebel differential is given by
\begin{equation}
\label{eq: strebel differential on P1}
q = -\frac{1}{4\pi^2}\left( a \left(\frac{dx}{x}\right)^2 +
b \left(\frac{dx}{1-x}\right)^2 +
c \left(\frac{dx}{x(1-x)}\right)^2 \right),
\end{equation}
where
\begin{equation*}
\begin{cases}
a = \frac{1}{2} \left( a_0 ^2 + a_{\infty} ^2 - a_1 ^2\right)\\
b = \frac{1}{2} \left( a_1 ^2 + a_{\infty} ^2 - a_0 ^2\right)\\
c = \frac{1}{2} \left( a_0 ^2 + a_1 ^2 - a_{\infty} ^2\right) .
\end{cases}
\end{equation*}
The behavior of the foliation of the
Strebel differential $q$ depends on the \emph{discriminant}
$$
ab + bc + ca =
\frac{1}{4} \left(a_0 + a_1 + a_{\infty} \right)
\left( a_0  + a_{\infty}  - a_1 \right)
\left( a_1  + a_{\infty}  - a_0 \right)
\left( a_0  + a_1  - a_{\infty} \right).
$$

\begin{Case} $ab + bc + ca >0$. The graph
is trivalent with two vertices and three edges,
as given in Figure~\ref{fig: case1}.
The two vertices
are located at
\begin{equation}\label{eq: roots}
\frac{a \pm i\sqrt{ab + bc + ca}}{a+b} ,
\end{equation}
and the length of edges $L_1$, $L_2$ and $L_3$ are given by
\begin{equation}\label{eq: cace1 length}
\begin{cases}
L_1 = \frac{1}{2} \left( a_0  + a_{\infty}  - a_1 \right)=
\frac{1}{2} \left( \sqrt{a + c}  + \sqrt{a + b}  - \sqrt{b + c}\right)\\
L_2 = \frac{1}{2} \left( a_1  + a_{\infty}  - a_0 \right)=
\frac{1}{2} \left( \sqrt{b + c}  + \sqrt{a + b}  - \sqrt{a + c} \right)\\
L_3 = \frac{1}{2} \left( a_0  + a_1  - a_{\infty} \right)=
\frac{1}{2} \left(\sqrt{a + c}  + \sqrt{b + c}  - \sqrt{a + b} \right) .
\end{cases}
\end{equation}
Note that positivity of $L_1,L_2$ and $L_3$ follows
from $ab + bc + ca >0$.
The
space of metric ribbon graphs with ordered boundary
in this case is
just $\mathbb{R}_+ ^3$ because there is only one
ribbon graph with boundary order
of this type and the only graph automorphism
that preserves the boundary
order is  the identity transformation.

\begin{figure}[htb]
\centerline{\epsfig{file=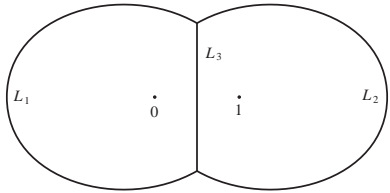, width=2in}}
\caption{The critical horizontal
leaves of the Strebel differential for
$ab + bc + ca >0$.}\label{fig: case1}
\end{figure}

The natural $\mathfrak{S}_3$-action on
the space of $(a_0,  a_1,   a_{\infty})$
 induces   faithful permutations of $L_1$,
$L_2$ and $L_3$ through (\ref{eq: cace1 length}).
The geometric picture  can be easily seen
from Figure~\ref{fig: case1a}. The normal subgroup
$\mathbb{Z}/3\mathbb{Z}$
of $\mathfrak{S}_3$ acts on $\mathbb{P}^1$ as
rotations about the axis connecting the north pole and
the south pole, where the poles of Figure~\ref{fig: case1a}
represent the zeroes (\ref{eq: roots}) of the Strebel
differential. Note that the three non-compact
leaves intersect at a zero of $q$ with $120^{\circ}$
angles. The action of the whole group $\mathfrak{S}_3$
is the same as
the dihedral group $D_3$ action on the triangle
$\triangle 01\infty$. As a result, $\mathfrak{S}_3$
acts faithfully on $(L_1,L_2,L_3)$ as its group of
permutations.
\end{Case}

\begin{figure}[htb]
\centerline{\epsfig{file=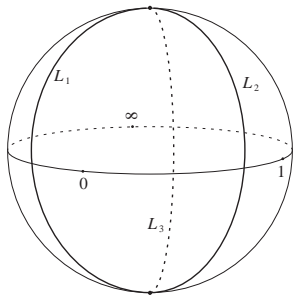, width=1.5in}}
\caption{A degree $3$ graph on a sphere.}\label{fig: case1a}
\end{figure}

The special case  $(a_0,a_1,a_\infty)
= (2,2,2)$ is of particular interest. The
Strebel differential (\ref{eq: strebel differential on P1})
is the pull-back of the building block
$q_0$ of (\ref{eq: building block}) via a rational
map
\begin{equation}
\label{eq: rational map of P1 to P1}
\zeta = \frac{4(x^2-x+1)^3}{27x^2(1-x)^2},
\end{equation}
and the ribbon graph Figure~\ref{fig: case1}
is the inverse image of the interval $[0,1]$ of this
map.

\begin{Case} $ab + bc + ca =0$. There are
three  ribbon graphs with labeled boundary components
 in this case,  whose
underlying graph has 1 vertex of
 degree $4$  and two edges
(Figure~\ref{fig: case2}). The vertex is located at
$a/(a+b)$.
Each of the three graphs corresponds
to one of the three factors,
$\left( a_0  + a_{\infty}  - a_1 \right)$,
$\left( a_1  + a_{\infty}  - a_0 \right)$, and
$\left( a_0  + a_1  - a_{\infty} \right)$, of the discriminant
being equal to $0$. For example, when
$\left( a_0  + a_1  - a_{\infty} \right) = 0$,
and
the lengths of the edges are given by
\begin{equation*}
\begin{cases}
L_1 =  a_0  = \sqrt{a + c} \\
L_2  =  a_1 = \sqrt{b + c} \\
L_3 = 0 .
\end{cases}
\end{equation*}
The $\mathfrak{S}_3$-action on $( a_0 ,a_1,a_{\infty})$
interchanges the three types of  ribbon graphs with
boundary order
in Case~2. The automorphism group of the ribbon graph
of Case~2  is $\mathbb{Z}/2\mathbb{Z}$, and only the
identity element preserves the boundary order.
\end{Case}

\begin{figure}[htb]
\centerline{\epsfig{file=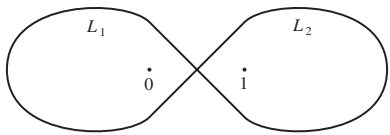, width=2in}}
\caption{Case $ab + bc + ca =0$.}\label{fig: case2}
\end{figure}

\begin{Case} $ab + bc + ca <0$. The underlying graph
is  of degree $3$ with two vertices and three edges,
but the topological type is different
from Case~1 (Figure~\ref{fig: case3}).
The two vertices are on the real axis located at
$$
\frac{a \pm \sqrt{-(ab + bc + ca)}}{a+b} .
$$
There are again three different  ribbon graphs
with ordered boundary,
each of which corresponds to one of the three factors
of the discriminant being negative. For example,
if $\left( a_0  + a_1  - a_{\infty} \right) < 0$, then
the length of edges are given by
\begin{equation*}
\begin{cases}
L_1 = a_0 =\sqrt{a + c}  \\
L_2 =  a_1  = \sqrt{b + c} \\
L_3 = \frac{1}{2} \left(- a_0  - a_1  + a_{\infty} \right)=
\frac{1}{2} \left(-\sqrt{a + c}  - \sqrt{b + c}  + \sqrt{a + b} \right) .
\end{cases}
\end{equation*}
$L_3$ is positive because $ab + bc + ca <0$.
\end{Case}

\begin{figure}[htb]
\centerline{\epsfig{file=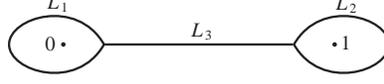, width=2in}}
\caption{Case $ab + bc + ca <0$.}\label{fig: case3}
\end{figure}

In Case~2 and Case~3, the automorphism group of the
 ribbon graph without ordered boundaries
is $\mathbb{Z}/2\mathbb{Z}$.
In every case, we can make the length of edges
arbitrary by a suitable choice of $(a_0, a_1, a_{\infty})$.
The discriminant
$ab+bc+ca$ divides the space $\mathbb{R}_+ ^3$
of triples $(a_0, a_1, a_{\infty})$
into 7 pieces: 3 copies of $\mathbb{R}_+ ^3$
along the $a_0$, $a_1$ and $a_{\infty}$ axes
where the discriminant is negative,
the center piece of $\mathbb{R}_+ ^3$ characterized
by positivity of the discriminant, and 3 copies of
$\mathbb{R}_+ ^2$ separating the 4 chambers that
correspond to the zero points of the discriminant
(Figure~\ref{fig: 03partition}):
\begin{equation}\label{eq: RG03 with lable}
\mathbb{R}_+ ^3
= \mathbb{R}_+ ^3 \coprod
\mathbb{R}_+ ^2\coprod
\mathbb{R}_+ ^2\coprod
\mathbb{R}_+ ^2\coprod
\mathbb{R}_+ ^3\coprod
\mathbb{R}_+ ^3\coprod
\mathbb{R}_+ ^3  .
\end{equation}

The product group $\mathbb{R}_+ ^3$ acts on
the space of $(a_0,a_1,a_{\infty})$ by multiplication,
but the action does not preserve the canonical
rational cell-decomposition of $RGB_{0,3} ^{\rm{met}}$.
Indeed, this action changes the sign of
the discriminant.

The three $\mathbb{R}_+ ^3$'s along the axes are
equivalent under the $\mathfrak{S}_3$-action
on the space of $(a_0, a_1, a_{\infty})$, and
each has a $\mathfrak{S}_2$-symmetry.
The three walls separating the chambers
are also equivalent
under the $\mathfrak{S}_3$-action, and again have the
same symmetry. Only the central chamber
is acted on by the full $\mathfrak{S}_3$.
Thus we have
\begin{equation}\label{eq: case03}
\begin{CD}
\mathbb{R}_+ ^3
@= \mathbb{R}_+ ^3 \coprod
\mathbb{R}_+ ^2\coprod
\mathbb{R}_+ ^2\coprod
\mathbb{R}_+ ^2\coprod
\mathbb{R}_+ ^3\coprod
\mathbb{R}_+ ^3\coprod
\mathbb{R}_+ ^3 \\
@VVV @VVV\\
\mathbb{R}_+ ^3/\mathfrak{S}_3
@= \mathbb{R}_+ ^3/\mathfrak{S}_3 \coprod
\mathbb{R}_+ ^2/\mathfrak{S}_2 \coprod
\mathbb{R}_+ ^3/\mathfrak{S}_2 .
\end{CD}
\end{equation}
\end{ex}

\begin{figure}[htb]
\centerline{\epsfig{file=ribbonfig/fig03partitionnew.eps, width=2in}}
\caption{Partition of $RGB_{0,3} ^{\rm{met}}=
\mathbb{R}_+ ^3$. }\label{fig: 03partition}
\end{figure}

The multiplicative group $\mathbb{R}_+$ acts naturally
on the ribbon graph complexes
$RG_{g,n} ^{\rm{met}}$ and $RGB_{g,n} ^{\rm{met}}$
by the multiplication of all the edge lengths by a constant.
Since the graph automorphism groups
$\Aut(\Gamma)$ and $\Aut_{\partial}(\Gamma)$
act on the edge space $\mathbb{R}_+ ^{e(\Gamma)}$
through a permutation of coordinate axes, the
multiplicative $\mathbb{R}_+$-action and the action of the
graph automorphism groups commute.
 Therefore, we have well-defined quotient complexes
$RG_{g,n} ^{\rm{met}}/\mathbb{R}_+$
and $RGB_{g,n} ^{\rm{met}}/\mathbb{R}_+$.
Since $\mathbb{R}_+ ^{e(\Gamma)}$ is a cone over the
$(e(\Gamma)-1)$-dimensional regular
$e(\Gamma)$-hyperhedron
$\Delta(123\cdots e(\Gamma))$
and since the graph automorphism groups act on
the hyperhedron,
the quotient of each rational cell is a rational
simplex
$$
\frac{\Delta(123\cdots e(\Gamma))}
{\Aut(\Gamma)}.
$$
Thus the quotient complexes
$RG_{g,n} ^{\rm{met}}/\mathbb{R}_+$
and $RGB_{g,n} ^{\rm{met}}/\mathbb{R}_+$
are rational simplicial complexes.

These quotient complexes are orbifolds modeled
on
$$
\frac{X_{\succeq\Gamma} ^{\rm{met}}/
\mathbb{R}_+}
{G},
$$
where $G$ denotes either $\Aut(\Gamma)$ or
$\Aut_{\partial}(\Gamma)$.
{F}rom (\ref{eq: space of expansions}), we have
\begin{equation}
\label{eq: local quotient}
\frac{X_{\succeq\Gamma} ^{\rm{met}}/
\mathbb{R}_+}{\Aut(\Gamma)}
=
\frac{\Delta(123\cdots e(\Gamma))\times
\mathbb{R}^{\rm{codim}(\Gamma)}}
{\Aut(\Gamma)}.
\end{equation}
Since $\Delta(123\cdots e(\Gamma))$ is homeomorphic
to $\mathbb{R}_+ ^{e(\Gamma) -1}$, the
quotient complexes are topological orbifolds.

On the moduli space
$\mathfrak{M}_{g,n}\times \mathbb{R}_+ ^n$, the
multiplicative group $\mathbb{R}_+$
acts on the space of $n$-tuples $\mathbb{R}_+ ^n$
through the multiplication of constants. The action
has no effect on $\mathfrak{M}_{g,n}$. Thus
we have the quotient space
$$
\frac{
\mathfrak{M}_{g,n}\times \mathbb{R}_+ ^n}
{\mathbb{R}_+}
=\mathfrak{M}_{g,n}\times
\Delta(123\cdots n).
$$
The bijection of Theorem~\ref{thm: Mgn = RGgn}
is equivariant under the $\mathbb{R}_+$-action,
and we have
\begin{equation}
\label{eq: bijection of the quotients}
\mathfrak{M}_{g,n}\times
\Delta(123\cdots n)
= \coprod_{\Gamma\in RGB_{g,n}}
\frac{\Delta(123\cdots e(\Gamma))}{\Aut_{\partial}
(\Gamma)}.
\end{equation}
This gives us an orbifold realization of
the space $\mathfrak{M}_{g,n}\times
\Delta(123\cdots n)$ as a rational simplicial complex.
When $n=1$, the space $\Delta(1)$ consists of just a point.
Therefore, we have a rational simplicial complex
realization
\begin{equation}
\label{eq: Mg1 as an orbifold}
\mathfrak{M}_{g,1} = \coprod_{\Gamma\in RG_{g,1}}
\frac{\Delta(123\cdots e(\Gamma))}{\Aut(\Gamma)}.
\end{equation}

\section{Belyi maps and algebraic curves defined
over $\overline{\mathbb{Q}}$}
\label{sect: belyi}

We have shown that a metric ribbon graph
defines a Riemann surface and
a Strebel differential on it. One can ask a
question: when does this Riemann surface
have the structure of an algebraic curve
defined over $\overline{\mathbb{Q}}$?
Using Belyi's theorem \cite{Belyi},
we can answer this question.

\begin{Def}
\label{def: belyi map}
Let $C$ be a nonsingular Riemann surface.
A \emph{Belyi map} is a holomorphic map
$$
f:C\longrightarrow \mathbb{P}^1
$$
that is ramified only at $0$, $1$ and $\infty$.
\end{Def}

\begin{thm}[Belyi's Theorem \cite{Belyi}]
\label{thm: belyi}
A nonsingular Riemann surface $C$ has the
structure of an algebraic curve defined
over $\overline{\mathbb{Q}}$ if and only if
there is a Belyi map onto $\mathbb{P}^1$.
\end{thm}

\begin{figure}[htb]
\centerline{\epsfig{file=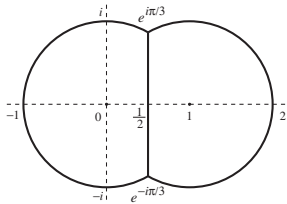, width=1.5in}}
\caption{The inverse image of $[0,1]$
via $\zeta = \frac{4(x^2-x+1)^3}{27x^2(1-x)^2}$.
It is obtained by joining two circles of radius $1$
at $1/2$.}
\label{fig: belyi03}
\end{figure}

\begin{cor}
\label{cor: trivalent belyi map}
A nonsingular Riemann surface $C$ has the
structure of an algebraic curve defined
over $\overline{\mathbb{Q}}$ if and only if
there is a Belyi map
$$
f: C\longrightarrow \mathbb{P}^1
$$
such that the ramification degrees over
$0$ and $1$ are $3$ and $2$, respectively.
Such a Belyi map is called \emph{trivalent}.
\end{cor}
\begin{proof}
Let $C$ be an algebraic curve over
$\overline{\mathbb{Q}}$ and
$h:C\longrightarrow \mathbb{P}^1$ a
Belyi map. Then the composition  $f=\phi\circ h$ of $h$ and
$$
\phi: \mathbb{P}^1\owns x\longmapsto
\zeta = \frac{4(x^2-x+1)^3}{27x^2(1-x)^2}
\in \mathbb{P}^1
$$
of (\ref{eq: rational map of P1 to P1})
gives a trivalent Belyi map. We first note that
$$
\frac{d\zeta}{dx} =
-\frac{4(x-2)(x+1)(2x-1)(x^2-x+1)^2}
{27x^3(1-x)^3}.
$$
Thus the ramification points of $\phi$ are
$x=-1$, $x=1/2$, $x=2$, and
$$
x = \frac{1\pm i\sqrt{3}}{2}.
$$
The critical values of $\phi$ are
$$
\phi(-1) = \phi(1/2) = \phi(2) = 1
$$
and
$$
\phi\left(\frac{1\pm i\sqrt{3}}{2}\right) = 0,
$$
and the ramification degrees at $1$ and $0$ are
$2$ and $3$, respectively.
$\phi$ sends $\{0,1,\infty\}$ to $\infty$, at which
it is also ramified. Since $h$ is not ramified
at $e^{\pm i\pi/3}$, $-1$, $1/2$, or $2$, the
composed map $f$ is ramified only at $0$, $1$,
and $\infty$ with the desired ramification degrees.

The inverse image of the
interval $[0,1]$ via $\phi$ is a ribbon graph of
Figure~\ref{fig: belyi03}. This graph is obtained
by adjoining two circles of radius $1$
that are centered at
$0$ and $1$ together with a common boundary at
$1/2$.
\end{proof}

\begin{Def}
\label{def: child's drawing}
A \emph{child's drawing}, or Grothendieck's
\emph{dessin d'enfant}, is
the inverse image of the line segment
$[0,1]$ by a Belyi map.
\end{Def}

\begin{thm}
\label{thm: riemann surface over Q-bar}
Let $\Gamma_{\rm{met}}$ be a metric ribbon
graph with no vertices of degree less than $3$.
It gives rise to an algebraic curve defined
over $\overline{\mathbb{Q}}$ if
all the edges
have the same length, which can be chosen to be
$1$. The metric ribbon graph determines a
unique Belyi map
$$
f: C(\Gamma_{\rm{met}})\longrightarrow \mathbb{P}^1
$$
such that the child's drawing associated with $f$
is the edge refinement of
$\Gamma_ {\rm{met}}$. The Strebel differential $q$
on $C(\Gamma_{\rm{met}})$ is the pull-back of the
building block $q_0$ of (\ref{eq: building block})
via the Belyi map,
i.e.,
$$
q = f^*\left(\frac{(d\zeta)^2}{4\pi^2\zeta(1-\zeta)}
\right).
$$

Conversely, every nonsingular algebraic curve
over $\overline{\mathbb{Q}}$ can be constructed
from a trivalent metric ribbon graph with edge length
$1$.
\end{thm}

\begin{proof}
Let $\Gamma_{\rm{met}}$
be a metric ribbon graph whose edges
all have length $1$,
$C(\Gamma_{\rm{met}})$ be the Riemann
surface defined by the metric ribbon graph,
 and $q$ the Strebel
differential on $C(\Gamma_{\rm{met}})$
whose noncompact leaves are $\Gamma_{\rm{met}}$.
We take the canonical triangulation
$\Delta_q$ of
Theorem~\ref{thm: canonical triangulation of a Riemann surface}
and the canonical coordinate system of
Definition~\ref{def: canonical coordinate system}.

For every edge $E$ of $\Gamma_{\rm{met}}$,
we define a map from a triangle with base $E$
into $\mathbb{P}^1$ as follows.
The triangle can be identified with the upper half
of the strip of Figure~\ref{fig: strip2}. So we define
\begin{equation}
\label{eq: map from strip to P1}
\{z\in \mathbb{C}\;|\;0\le Re(z)\le 1, Im(z)\ge 0\}
\owns z\longmapsto \zeta=
\sin^2(\pi z)\in \mathbb{P}^1.
\end{equation}
We note that (\ref{eq: map from strip to P1})
is equivalent to
\begin{equation}
\label{eq: dz2 = q0}
(dz)^2 = q_0 =
\frac{1}{4\pi^2}\frac{(d\zeta)^2}{\zeta(1-\zeta)}.
\end{equation}

We wish to show that this map consistently
extends to a holomorphic Belyi map
$$
f: C(\Gamma_{\rm{met}})\longrightarrow \mathbb{P}^1.
$$

The  map (\ref{eq: map from strip to P1})
extends to the whole strip in an obvious way.
The point $z=1/2$ is mapped to $\zeta = 1$, at
which the map is ramified with ramification
degree $2$.

At a vertex $V$ to which $E$ is incident, the canonical
coordinate is given by
$$
w = 2^{2\pi i k/m} z^{2/m}
$$
as in (\ref{eq: zeroglue}), where
$m$ is the degree of $V$ and $k$ is an integer.
In terms of the $w$-coordinate, the map
(\ref{eq: map from strip to P1}) is given by
$$
\zeta = \sin^2(\pm \pi w^{m/2})
= \pi^2 w^m + \cdots.
$$
This expression does not depend on the choice of
an edge attached to $V$, hence the map
(\ref{eq: map from strip to P1}) extends consistently
to a neighborhood of $V$. The map is ramified
at $\zeta = 0$ with local ramification degree $m$.

Finally, let us consider a boundary component of
$\Gamma_{\rm{met}}$ consisting of $k$ edges.
{F}rom (\ref{eq: poleglue}), we have a local coordinate
$u$ on the boundary disk that is given by
$$
u = \exp\left(\frac{2\pi i(j+z)}{k}\right),
$$
where $j$ is an integer. Then
$$
z=\frac{k}{2\pi i} \log u -j.
$$
Noting that
$$
\sin^2(x) = -\frac{e^{2i x} + e^{-2ix} -2}{4},
$$
we have
\begin{equation}
\label{eq: belyi map u to P1}
\zeta = -\frac{1}{4}
(u^k + u^{-k} -2).
\end{equation}
This map sends $u=0$ to $\zeta = \infty$, and
is independent of the choice of edge around
$u=0$ and  branch of the logarithm function.
The map (\ref{eq: belyi map u to P1}) is equivalent
to the relation
$$
-\frac{k^2}{4\pi^2} \frac{(du)^2}{u^2}
= \frac{(d\zeta)^2}{4\pi^2\zeta(1-\zeta)} = q_0.
$$
We have thus shown that the map
(\ref{eq: map from strip to P1}) extends to a
holomorphic map $f$
from
the whole Riemann surface $C(\Gamma_{\rm{met}})$
 onto
$\mathbb{P}^1$ that is ramified only at
$0,1,\infty$. The unique Strebel differential $q$
is given by $f^*q_0$.

Conversely, let $C$ be a nonsingular algebraic curve
defined over
$\overline{\mathbb{Q}}$. Let
$$
f:C\longrightarrow \mathbb{P}^1
$$
be a trivalent Belyi map. Then the inverse
image of the interval $[0,1]$ via $f$ is
the edge refinement of a trivalent
metric ribbon graph on $C$ whose edge length is
$1$ everywhere. It is the union of noncompact
leaves of the Strebel differential
$f^*q_0$ on $C$. Starting from this child's drawing,
we recover the complex structure of $C$.
\end{proof}

We have already given two examples of genus $1$ in
Section~\ref{sect: strebel} and one
example of genus $0$ in Section~\ref{sect: combi}.

Theorem \ref{thm: riemann surface over Q-bar}
does not completely characterize which metric ribbon graphs
correspond to algebraic curves defined over
$\overline{\mathbb{Q}}$. Furthermore, in Grothendieck's {\em dessin
d'enfant} graphs which have vertices of degree 2 and 1 also
appear. It is possible to incorporate the vertices of degree 2 coming
from these child's drawings in terms of usual ribbon graphs (those
with no vertices having degree less than 3)
by sharpening the statement of the theorem as
follows.
\begin{cor}
\label{cor: degree 2 graphs}
Let $\Gamma_{\rm{met}}$ be a metric ribbon graph, such that the
ratios of the lengths of its edges are all rational, (so they can
be chosen to be positive integers). Then there is a unique Belyi
map $f:C(\Gamma_{\rm{met}})\longrightarrow \mathbb{P}^1$, such that
if $\tilde\Gamma_{\rm{met}}$ is the metric graph obtained by
replacing each edge of length $n$ with $n$ edges of length 1 by inserting $n-1$
vertices of degree 2 in the edge, then the child's
drawing associated with $f$ is the edge refinement of
$\tilde\Gamma_{\rm{met}}$. In addition, the Strebel differential $q$
on $C(\Gamma_{\rm{met}})$ is given by $q=f^*q_0$.
\end{cor}
\begin{proof}
Essentially, the only change in the proof of the theorem needed is
to replace the map $\sin^2(\pi z)$ from the strip associated with
an edge $E$ with the map $\sin^2(n\pi z)$. This will add $n-1$
additional zeros on the edge, each with ramification degree $2$.
\end{proof}

To obtain the complete classification of metric ribbon graphs
which correspond to algebraic curves defined over
$\overline{\mathbb{Q}}$, we shall have to consider ribbon graphs
which have vertices of degree 1. In the theory of Strebel
differentials, vertices of degree 1 correspond to poles of order 1 of the
quadratic differential.
There is a uniqueness theorem concerning
Strebel differentials which have poles of order at most 2 as well.
See Theorem 7.6 in \cite{Looijenga} for a precise statement of the
result.

For our purposes, we really only need to address the question of how to
construct a Riemann surface corresponding to a metric ribbon
graph. The reader can easily verify that the construction of a Riemann surface
corresponding to a metric ribbon graph
given in Theorem \ref{thm: Mgn = RGgn}
still applies when we allow the graphs to have vertices of
degree 1 or 2.  Furthermore, the construction also yields a
quadratic differential, which has poles of order 1 for vertices of
degree 1, and neither a pole nor a zero for vertices of degree 2
(although they will lie on the critical trajectories).

The same methods as in Theorem \ref{thm: riemann surface over Q-bar}
allow one to construct a Belyi map from the Riemann surface
corresponding to this more general type of ribbon graph, by associating the metric
ribbon graph with all edges having length 1 to the graph. Putting
this all together, we come up with the following.
\begin{thm}
\label{thm: bijective correspondence}
There is a one to one correspondence between the following:
\begin{enumerate}
\item The set of isomorphism classes of ribbon graphs.
\item The set of isomorphism classes of child's drawings.
\item The set of isomorphism classes of Belyi maps.
\end{enumerate}
This correspondence is given as follows. A ribbon graph $\Gamma$
corresponds to a metric ribbon graph $\Gamma_{{\rm met}}$
with all edges having length
1, giving rise to a Riemann surface $C(\Gamma_{{\rm met}})$ and a
Strebel differential $q$ which is the pullback of the quadratic
differential $q_0$ by the unique Belyi map from $C(\Gamma_{{\rm met}})$
to $\mathbb{P}^1$ whose associated child's drawing is the edge
refinement of $\Gamma_{\rm met}$. Moreover, this correspondence between
ribbon graphs and Belyi maps agrees with the Grothendieck
correspondence.
\end{thm}

Grothendieck's use of the terminology {\em child's drawing} to
illustrate the relationship between graphs and algebraic curves
emphasizes how strange and beautiful it is that a deep area of
mathematics can be described in such simple terms. In our
construction, we have shown how to take a Child's drawing,
associate a metric ribbon graph to it, construct a Riemann surface
equipped with a quadratic differential, as well as a Belyi map
from this surface to $\mathbb{P}^1$. It is amazing how much
information is concealed within such a simple picture.

\providecommand{\bysame}{\leavevmode\hbox to3em{\hrulefill}\thinspace}


\begin{thebibliography}{10}

\bibitem{Belyi}
G.~V.~Belyi, \emph{On galois extensions of a maximal cyclotomic fields}, Math.\
  U.S.S.R.\ Izvestija \textbf{14} (1980), 247--256.

\bibitem{Gardiner}
Fredderick~P. Gardiner, \emph{{Teichm\"uller} theory and quadratic
  differentials}, John Wiley \& Sons, 1987.

\bibitem{Harer}
John~L. Harer, \emph{The cohomology of the moduli space of curves}, 
in Theory of
  Moduli, Montecatini Terme, 1985 (Edoardo Sernesi, ed.), Springer-Verlag,
  1988, pp.~138--221.

\bibitem{Harer-Zagier}
John~L. Harer and Don Zagier, \emph{The {Euler} characteristic of the moduli
  space of curves}, Inventiones Mathematicae \textbf{85} (1986), 457--485.

\bibitem{Looijenga}
Eduard Looijenga, \emph{Cellular decompositions of compactified moduli spaces
  of pointed curves}, in Moduli Space of Curves (R.~H.~Dijkgraaf et~al., ed.),
  Birkhaeuser, 1995, pp.~369--400.

\bibitem{Mulase1995}
Motohico Mulase, \emph{Asymptotic analysis of a hermitian matrix integral},
  International Journal of Mathematics \textbf{6} (1995), 881--892.

\bibitem{Penner}
Robert~C. Penner, \emph{Perturbation series and the moduli space of {Riemann}
  surfaces}, Journal of Differential Geometry \textbf{27} (1988), 35--53.

\bibitem{Sansone-Gerretsen}
Giovanni Sansone and Johan Gerretsen, \emph{Lectures on the theory of functions
  of a complex variable, volume i and ii}, Wolters-Noordhoff Publishing, 1960,
  1969.

\bibitem{Satake}
Ichiro Satake, \emph{The {Gauss-Bonnet} theorem for {V}-manifold}, Journal of
  the Mathematical Society of Japan \textbf{9} (1957), 464--492.

\bibitem{Schneps}
Leila Schneps, \emph{The grothendieck theory of dessins d'enfants}, 
  London Mathematical Society Lecture Notes Series,
vol.\ 200, 1994.

\bibitem{STT}
Daniel~D. Sleator, Robert~E. Tarjan, and William~P. Thurston, \emph{Rotation
  distance, triangulations, and hyperbolic geometry}, Journal of the American
  Mathematical Society \textbf{1} (1988), 647--681.

\bibitem{Strebel}
Kurt Strebel, \emph{Quadratic differentials}, Springer-Verlag, 1984.

\bibitem{Thurston}
William Thurston, \emph{Three-dimensional geometry and topology, volume 1 and
  2}, Princeton University Press, 1997, (volume 2 to be published).

\end{thebibliography}
\bibliographystyle{amsplain}

\end{document}